\documentclass[aps,pra,groupedaddress,twocolumn,notitlepage,superscriptaddress,10pt]{revtex4-1}
\usepackage{epsfig}

\usepackage{amsfonts}

\usepackage{graphicx}
\usepackage{pst-all}	%call the pstricks package
\usepackage{braket}
\usepackage{amsmath}
\usepackage{amssymb}
\usepackage{xcolor}
\usepackage{bbm}
\usepackage{revsymb}
\usepackage{yfonts}
\usepackage{tensor} % for pre subscript

\definecolor{lightgray}{rgb}{0.925,0.925,0.925}
\definecolor{lightyellow}{rgb}{1,0.925,0.58}
\definecolor{darkblue}{rgb}{0.1,0.2,0.6}
\definecolor{darkred}{rgb}{0.8,0.1,0.2}
\usepackage[colorlinks,citecolor=darkblue,linkcolor=darkred,urlcolor=darkblue,breaklinks]{hyperref} 
\usepackage{tikz}
\usepackage{enumitem}
\setlist[enumerate]{itemsep=0mm}
\usepackage{setspace}

\newcommand{\defineas}{\mathrel{\overset{\textrm{def}}{=}}}
\newcommand{\bigket}[1]{ {\big|{#1}\big\rangle} }

\newcommand{\bigbra}[1]{ {\big\langle{#1}\big|} }

\renewcommand{\Tr}{\operatorname{Tr}}

\newcommand{\Hi}{{\cal H}}

\newcommand{\Z}{\mathbb{Z}}

\newcommand{\rT}{{\rho^{T_A}} }
\newcommand{\norm}[1]{\left\lVert#1\right\rVert}
\newcommand{\aTr}{\operatorname{\widetilde{Tr}}}
\newcommand{\arho}{\tilde{\rho}}

\numberwithin{equation}{section}
\renewcommand\theequation{\arabic{section}.\arabic{equation}}

\usepackage{amsthm}
\theoremstyle{plain}
\newtheorem{thm}{Theorem}
\newtheorem{corr}{Corollary}
\newtheorem{conjecture}{Conjecture}
\newtheorem*{conjecture*}{Conjecture}

\theoremstyle{definition}

\begin{document}	
\title{{\bf Anyonic Partial Transpose I:
Quantum Information Aspects}}
%\author{***}

\author{Hassan Shapourian%
%  \thanks{Electronic address: \texttt{hassan.shapp@gmail.com}}
  }
\affiliation{Microsoft Station Q, Santa Barbara, CA 93106, USA}
\affiliation{Department of Physics, Harvard University, Cambridge, MA~02138}
\affiliation{Department of Physics, Massachusetts Institute of Technology,
Cambridge, MA~02139}

\author{Roger S.~K.~Mong}
\affiliation{Department of Physics and Astronomy and Pittsburgh Quantum Institute, University of Pittsburgh, Pittsburgh, PA~15260}

\author{Shinsei Ryu}
\affiliation{ Department of Physics, Princeton University, Princeton, New Jersey, 08540, USA}

\date{\today}

\begin{abstract}
A basic diagnostic of entanglement in mixed quantum states is known as the partial transpose and the corresponding entanglement measure is called the logarithmic negativity. Despite the great success of logarithmic negativity in characterizing bosonic many-body systems, generalizing the partial transpose to fermionic systems remained a technical challenge until recently when a new definition that accounts for the Fermi statistics was put forward. In this paper, we propose a way to generalize the partial transpose to anyons with (non-Abelian) fractional statistics based on the apparent similarity between the partial transpose and the braiding operation. We then define the anyonic version of the logarithmic negativity and show that it satisfies the standard requirements such as monotonicity to be an entanglement measure. In particular, we elucidate the properties of the anyonic logarithmic negativity by computing it for a toy density matrix of a pair of anyons within various categories. We conjecture that the subspace of states with a vanishing logarithmic negativity is a set of measure zero in the entire space of anyonic states, in contrast with the ordinary qubit systems where this subspace occupies a finite volume. We prove this conjecture for multiplicity-free categories.
\end{abstract}

\maketitle
%{\hypersetup{linkcolor=black} \tableofcontents}

\section{Introduction}

Characterizing quantum systems based on their encoded entanglement and quantifying the amount of entanglement in terms of a computable measure is a basic yet fundamental question across various fields of physics research from quantum information to condensed matter~\cite{Amico_rev2008,Calabrese2004,*Calabrese2009,Levin_Wen2006,Kitaev_Preskill2006,WenChen_book} and high energy theory~\cite{HOLZHEY1994443,RyuTakayanagi,Rangamani_book}. 

Consider a product Hilbert space $\Hi_A\otimes\Hi_B$ of two parties $A$ and $B$. The bipartite entanglement of a pure state $\ket{\Psi}$ in such Hilbert space is measured in terms of von~Neumann entanglement entropy $S_1(\rho_A)=-\Tr (\rho_A\ln\rho_A)$ or R\'enyi entanglement entropies $S_n(\rho_A)=\ln \Tr (\rho_A^n)/(1-n)$, where $\rho_A=\Tr_B(\ket{\Psi}\bra{\Psi})$ is the reduced density matrix of subsystem $A$ after partially tracing over
$\Hi_B$ and $n$ is a positive number.
$S(\rho_A) = S(\rho_B)$ characterizes the amount quantum entanglement between the two parties.

Despite the importance of the von~Neumann entanglement entropy as a diagnosis for
many-body quantum pure states,
it fails to correctly quantify the bipartite entanglement in mixed states, when the state of the composite system is described by a density matrix $\rho$. Addressing the issue of quantifying mixed-state entanglement is not only of interest from a basic research point of view, for example when one wants to identify multi-partite entanglement, but also for practical purposes because of the ubiquity of mixed states in nature, i.e., we almost always deal with open quantum systems in laboratory.
The challenge is that even other entanglement quantities defined based on the von~Neumann entanglement entropy to quantify entanglement in mixed states such as the mutual information $I(A:B) = S_1(\rho_A) + S_1(\rho_B) - S_1(\rho_{AB})$ does not work since it does not distinguish between
quantum and classical correlations.
For instance, there exists a large family of classically correlated states called separable, while  their mutual information is non-vanishing.

A candidate test for quantum entanglement in mixed states is the \emph{partial transpose} (PT) criterion~\cite{Peres1996,Horodecki1996,Simon2000,PhysRevLett.86.3658,PhysRevLett.87.167904,Zyczkowski1,Zyczkowski2}, which is a diagnosis for non-separable states. The partially transposed density matrix 
can then be used to construct the logarithmic negativity (LN) as a measure of entanglement in mixed states~\cite{PlenioEisert1999,Vidal2002,Plenio2005}.
Logarithmic negativity has been shown to be useful in studying various many-body quantum systems including
harmonic oscillator chains~\cite{PhysRevA.66.042327,PhysRevLett.100.080502,PhysRevA.78.012335,Anders2008,PhysRevA.77.062102,PhysRevA.80.012325,Eisler_Neq,Sherman2016,dct-16}, quantum spin models~\cite{PhysRevA.80.010304,PhysRevLett.105.187204,*PhysRevB.81.064429,*PhysRevLett.109.066403,Ruggiero_1,PhysRevA.81.032311,PhysRevA.84.062307,Mbeng,Grover2018,java-2018,Gray2019,maccormack2020operator}, (1+1)d conformal and integrable field theories~\cite{Calabrese2012,*Calabrese2013,*Calabrese_Ft2015,Ruggiero_2,Alba_spectrum,kpp-14,fournier-2015,bc-16,Wald_2020,PhysRevB.101.064207,lu2019structure,angelramelli2020logarithmic,Juh_sz_2018,Schreiber,rosz2020entanglement,Shapourian_FS,wu2019entanglement,10.21468/SciPostPhys.8.4.063},  topologically ordered phases of matter~\cite{Wen2016_1,Wen2016_2,PhysRevA.88.042319,PhysRevA.88.042318,hc-18,PhysRevB.101.085136},
and in out-of-equilibrium dynamics \cite{ctc-14,ez-14,hb-15,ac-18b,wen-2015,gh-18,alba2020spreading,Gruber_2020,PhysRevB.101.245130,Kudler-Flam2019,*2020JHEP...04..074K,*2020arXiv200811266K,Grover2020}, 
 as well as holographic theories~\cite{Rangamani2014,2014JHEP...09..010K,PhysRevD.99.106014,PhysRevLett.123.131603} and variational~\cite{Clabrese_network2013,Alba2013,PhysRevB.90.064401,Nobili2015} and random states~\cite{Marko2007,Aubrun2010,*Aubrun2014,*Aubrun2012,Szyma_ski_2017,Collins2012,*Collins_rev,Shap_rand}. There are also experimental proposals to measure moments of the partially transposed density matrix with ion traps and cold atoms~\cite{Elben2019,gbbs-17,csg-18,Elben2020}.

In this paper, we take steps to propose a 
way to quantify mixed state entanglement in anyonic systems
where unlike the aforementioned (bosonic) systems the local operators do not necessarily commute. Anyons are quasi-particles with fractional statistics which are building blocks of a topological quantum computer~\cite{Nayak2008}.
In this regard, our proposal for an anyonic entanglement measure here can be useful in characterizing states in topological quantum computing.
 Our idea is based on a natural generalization of the PT to anyonic density matrices 
and
 inspired by the observation that PT may be viewed as a partial time-reversal transformation, where the arrow of time is reversed for one subsystem with respect to that of the other subsystem. 

 For general anyon models, we find that the process of reversing the arrow of time can be plausibly formulated in terms of a half-braid in the diagrammatic approach. 
We further demonstrate how such a construction works for several examples of entangled anyonic states.
As we will see, similar to the ordinary LN, the anyonic logarithmic negativity (ALN) captures exclusively anyonic correlations, i.e., it is an entanglement monotone and vanishes for separable states where there is no non-trivial charge line connecting the two subsystems.
These properties are clearly in contrast with the anyonic generalizations of von Neumann entropy which captures all correlations between the two subsystems. In this regard, the ALN shares some similarity with the anyonic charged entanglement entropy introduced in Ref.~\cite{KnappBonderson2017}.
Lastly, as a byproduct, we find that the phase factors appearing in the PT of Ising category lead to a sign factor identical to that of fermionic systems which was found in Ref.~\cite{Shap_pTR}.

The rest of this paper is organized as follows:
In Sec.\,\ref{sec: preliminaries}, we provide background materials about the partial transpose and anyonic states and entanglement.
In Sec.\,\ref{sec:Anyonic partial transpose}, 
we explain how PT can be implemented in the diagrammatic approach and derive an expression for the ALN of 
 an anyonic analogue of a dimer state.
Next in Sec.\,\ref{sec:properties}, we discuss various conditions which an entanglement measure of anyon models must satisfy and show that the proposed ALN fulfills them all.
In Sec.\,\ref{sec:Examples}, we explicitly calculate the ALN of the anyon dimer for the Fibonacci anyons, as well as some special cases of the $su(2)_k$, and $su(3)_k$ theories.
Finally, we finish our paper by several closing remarks on outstanding issues and new avenues for future research in Sec.\,\ref{sec:Conclusions}.
Some details of our calculations and background information are provided in five appendices.

%%%%%%%%%%%%%%%%%%%%%%%%%%%%%%%%%%%%%%%%%%%%%%%%%%
%%%%%%%%%%%%%%%%%%%%%%%%%%%%%%%%%%%%%%%%%%%%%%%%%%
\section{Preliminaries}
\label{sec: preliminaries}

This section is intended as a brief review of the partial transpose in ordinary bosonic and fermionic systems and the entanglement in anyonic systems.

\subsection{Review of bosonic and fermionic partial transpose}
\label{sec:review bosonic PT}

As mentioned earlier, separable states are classical states 
   which cannot be used to generate Bell pairs that are used in quantum key distribution or quantum teleportation protocols.
They take the general form 
\begin{align}
\label{eq:separable}
    \rho&=\sum_{i,j} p_{ij} \rho_A^{(i)} \otimes \rho_B^{(j)} && \text{with $p_{ij}\geq 0$},
\end{align}
where $\{ \rho_A^{(i)}  \}$ and $\{\rho_B^{(j)} \}$ are sets of local density matrices~\footnote{Here is a protocol to prepare a separable state by means of local operation and classical communication: Two parties $A$ and $B$ have a set of local density matrices $\{ \rho_A^{(i)} \}$ and $\{\rho_B^{(j)} \}$, respectively, and agree to prepare the $i$-th($j$-th) state when a classical random number generator outputs its $ij$-th outcome. The random number generator is designed such that it returns the $ij$-th output with probability $p_{ij}$.}. As these states are purely classical by construction, any faithful measure of entanglement must give zero when computed for them.

The PT of a state 
\begin{align}
\rho=\sum_{ijkl} \rho_{ijkl} \bigket{e_A^{(i)}, e_B^{(j)}}  \bigbra{e_A^{(k)}, e_B^{(l)}},    
\end{align}
written in a local orthonormal basis $\{\bigket{e_A^{(k)}}, \bigket{e_B^{(j)}}  \}$ is defined by exchanging the indices of subsystem $A$ (or $B$) as in
\begin{align} \label{eq:rTb}
\rho^{T_A}=\sum_{ijkl} \rho_{ijkl} \Ket{e_A^{(k)}, e_B^{(j)}}  \Bra{e_A^{(i)}, e_B^{(l)}}.
\end{align}
We note that $\rho^{T_A}$ is a Hermitian operator, and hence all its eigenvalues are real. The PT test is to check whether or not $\rho^{T_A}$ contains any negative eigenvalues.
A separable state by definition is not affected by the PT, i.e., it remains positive semi-definite even after PT.
On the other hand, a given state which has negative eigenvalues after PT (e.g.~Bell states) cannot be a separable state.  In this regard, PT test provides a necessary condition for separability~
    \footnote{However, the PT criterion is not a sufficient condition for separability. In other words, there exists a family of states which passes the PT test while they are not separable~\cite{Horodecki1997}.
    These states are also known as bound entangled where their entanglement cannot be distilled to carry out quantum computing processes such as teleportation~\cite{Horodecki1998}.
    The states which satisfy the PT test collectively form a convex set called the positive partial transpose (PPT) states.}.

The negative eigenvalues of the partially transposed density matrix can  be used to construct the logarithmic negativity (LN)~\cite{PlenioEisert1999,Vidal2002,Plenio2005},
\begin{align} 
    \label{eq:neg_def}
    {\cal E}(A:B) &= \ln \norm{\rT }_1,
\end{align}
where $\norm{X}_1= \Tr \sqrt{X X^\dag}$ is the trace norm (or one-norm). 
The PT is trace preserving, meaning that--although some eigenvalues are negative--the overall sum adds to one.
Since $\rT$ is Hermitian, the trace norm is simply the sum of the absolute value of  eigenvalues of $\rT$. Therefore, existence of negative eigenvalues implies a non-zero LN.

It is worth recapitulating the difference between the bosonic and fermionic PTs~\cite{Shiozaki_Ryu2017,Shap_unoriented,Shiozaki_antiunitary} at this point. Consider a bipartite Fock space of fermions where a state in this Hilbert space is denoted by $\bigket{\{ n_j \}_{j \in A} , \{n_j\}_{j \in B}}$
where $n_j=0,1$ are occupation numbers. Using this basis to represent density matrices, the transformation rule for the partial transpose is given by~\cite{Shap_pTR,Shiozaki_antiunitary},
\begin{align}
&\Big( \bigket{\{ n_j \}_{A} , \{n_j\}_{B}} \bigbra{\{ \bar n_j \}_{A}, \{ \bar n_j \}_{B} } \Big)^{T_A} \nonumber \\
&=(-1)^{\phi(\{n_j\}, \{\bar n_j\})}
 \bigket{\{ \bar n_j \}_{A} , \{n_j\}_{B}} \bigbra{\{ n_j \}_{A}, \{ \bar n_j \}_{B} }, \nonumber
%\label{eq:app_{ab}^f_21}
\end{align}
where the phase factor is
\begin{align}
\phi &= \frac{[(\tau_A+\bar{\tau}_A)\bmod2]}{2} + (\tau_A+\bar{\tau}_A)(\tau_B+\bar{\tau}_B),
\end{align}
and $\tau_s=\sum_{j\in s} n_j$,  and $\bar{\tau}_s=\sum_{j\in s} \bar{n}_j$ are the number of occupied modes in the segment $s=A,B$. 
%The complex phase factor 
The sign factor
turns out to be an essential difference between the bosonic PT (which is applied to spin chains, harmonic chains, qubits, etc.) and the fermionic PT, especially when it comes to partition functions of fermionic systems with a fixed spin structure~\cite{Shiozaki_antiunitary,Kobayashi}.
Although the fermionic PT can be derived as the only definition consistent with operator algebras in fermionic systems (see for instance Ref.~\cite{Shiozaki_antiunitary}),
there was 
little
physical intuition as to where this phase factor comes from.
 As we will see later in this paper, the sign factor can be reproduced if we define PT in terms of braiding of the underlying Majorana fermion wordlines. Before delving into details of anyon models, we refer an interested reader to 
Appendix~\ref{app:partial transpose and exchange}, where we uncover a close similarity between the fermionic PT of a density matrix of a Majorana dimer state and the exchange operator of two vortices in a $p_x+ip_y$ chiral superconductor.

%%%%%%%%%%%%%%%%%%%%%%%%%%%%%%%%%%%%%%%%%%%%%%%
%%%%%%%%%%%%%%%%%%%%%%%%%%%%%%%%%%%%%%%%%%%%%%%

\subsection{Anyonic state and entanglement}
\label{sec:Anyonic state and entanglement}

Our goal in this part is to adapt the standard notions of entanglement entropy, discussed in Introduction, to anyon models. 
We begin by reviewing some terminology regarding anyonic states (pure or mix) and the notion of entanglement in that context~\cite{Bonderson_thesis2012,BondersonShtengel2008,KnappBonderson2017}. 

Anyonic states and their algebraic properties can be defined axiomatically~\cite{BondersonShtengel2008,KnappBonderson2017}. A generic density matrix is a sum of projection operators using the fusion rules. As a result, the only input required for constructing anyonic density matrices is the associativity of fusion rules, i.e., $F$ symbols.
As we argue in this paper, a plausible choice to incorporate the PT in this formalism would be to reverse the arrow of time for anyon worldlines which are ultimately related to a set of braiding exchanges. 
We apply the proposed PT to various examples of anyonic density matrices and compare it with other candidate entanglement measures.
In the case of Ising anyons, we find that the anyonic PT and ALN reproduces the expected results identical to those of the entangled Majorana dimers~\cite{Shap_pTR}. 

Despite the fact that our derivation was carried out for some toy examples, the procedure of the anyonic PT is general and only assumes a normal ordering of anyon lines. Here, by normal ordering we mean that if one wants to apply PT to anyon lines in the middle of a diagram, those anyons must be brought to the edge of the diagram by braiding them past other anyons.
This is similar to the case of fermionic states where Fock space need to be normal ordered when taking the partial trace or partial transpose~\cite{Shap_sep}.

Throughout our discussion in this paper, we use the diagrammatic description of anyon states to implement the anyonic PT.
For clarity, we denote an anyonic state by $\arho$ and its associated diagrammatic trace (or quantum trace) by tilde. A generic state $\arho \in V^{a_1 \dots a_n}_{a_1 \dots a_n}$ 
%{\color{red}there are ${ }^{\prime}$ here but not in the picture below.}\HS{right. It must not be there.}
is shown by
\begin{align}
\arho \defineas
\label{eq:rho}
\psscalebox{.85}{
 \pspicture[shift=-1.](-1,-1.1)(1,1.2)
  \footnotesize
%%%%% Box:
  \psframe[linewidth=0.9pt,linecolor=black,border=0](-0.8,-0.5)(0.8,0.5)
  \rput[bl]{0}(-0.1,-0.1){\normalsize $\arho$}
  \rput[bl]{0}(-0.22,0.7){$\mathbf{\ldots}$}
  \rput[bl]{0}(-0.22,-0.75){$\mathbf{\ldots}$}
%%%%% Line connections:
  \psset{linewidth=0.9pt,linecolor=black,arrowscale=1.5,arrowinset=0.15}
  \rput[bl]{0}(0.5,1.1){$a_n$}
  \psline(0.6,0.5)(0.6,1)
  \rput[bl]{0}(-0.7,1.1){$a_1$}
  \psline(-0.6,0.5)(-0.6,1)
  \rput[bl]{0}(0.5,-1.25){$a_n$}
  \psline(0.6,-0.5)(0.6,-1)
  \rput[bl]{0}(-0.7,-1.25){$a_1$}
  \psline(-0.6,-0.5)(-0.6,-1)
%%%%% Arrows:
  \psline{->}(0.6,0.5)(0.6,0.9)
  \psline{->}(-0.6,0.5)(-0.6,0.9)
  \psline{-<}(0.6,-0.5)(0.6,-0.9)
  \psline{-<}(-0.6,-0.5)(-0.6,-0.9)
\endpspicture , 
}
\end{align}
which is a Hermitian semi-definite operator with unit quantum trace, $\aTr\arho=1$, where $\aTr$ is the quantum trace.
Appendix \ref{app:diagrams} reviews some basics of the tensor category, the so-called theory of anyons, such as the fusion rules, states, and operators following Refs.~\cite{Bonderson_thesis2012,BondersonShtengel2008,KnappBonderson2017}. 

Let us now review the definitions of several entanglement quantities in anyon models.
The anyonic von Neumann entanglement entropy or R\'enyi entropies are defined in terms of anyonic density matrices as 
\begin{align}
\label{eq:vN_ent}
S (\arho) &= -\aTr( \arho \ln \arho),\\
S_n(\arho) &= \frac{1}{1-n} \ln \aTr (\arho^n).
\end{align}
These quantities are collectively known as the anyonic entanglement entropies (AEE)~\cite{KnappBonderson2017}.
In practice, the anyonic von Neumann entropy of a given state is usually calculated by analytically continuing the R\'enyi entropies, that is
\begin{align}
S (\arho) = \lim_{n\to 1} S_n(\arho),
\end{align}
where $\aTr (\arho^n)$ is computed digarammatically.
Similarly, one can define
the mutual information
\begin{align} \label{eq:mutual}
I(A:B)=S(\arho_A)+S(\arho_B)-S(\arho_{AB}),
\end{align}
where $\arho_{AB}$ is the density of matrix of the bipartite system ($A \cup B$), 
while $\arho_A=\aTr_B(\arho_{AB})$ and $\arho_B=\aTr_A(\arho_{AB})$ are the corresponding reduced density matrices. Here, the partial tracing is performed in the anyonic sense as explained in Appendix \ref{app:diagrams}. Here also, the anyonic mutual information suffers from the same issue~\cite{KnappBonderson2017} that it overestimates the entanglement as in the case of ordinary qubits.
 
Another probe of the entanglement in anyonic systems (which may also be applicable to mixed states) is the entropy of anyonic charge entanglement (ACE)~\cite{KnappBonderson2017} which is defined by
\begin{align}
\label{eq:Sace}
S_{\text{ace}} (A:B) \defineas S\left( D_{A:B}[\arho]\right) - S\left( \arho\right),
\end{align}
where $D_{A:B}$ is the charge line decoherence superoperator that projects out the charge lines which connect the subsystems $A$ and $B$ unless it is the identity charge line (or equivalently no line).
 $D_{A:B}$ acts on 
a state
% the system 
 by applying the $\omega_0$-loop~\cite{Bonderson2009,KnappBonderson2017}:
\begin{align}
  \begin{pspicture}[shift=-1.05](1.1,-3.3)(2.8,-1.2)
        \scriptsize
        \psellipse[linecolor=black,border=0](1.9,-2.2)(0.2,.6) \rput(1.45,-2.35){$\omega_0$}
        \psline{>-}(1.71,-2.25)(1.71,-2.15)
        \psline[border=2pt](2.6,-1.2)(1.9,-1.9)
        \psline(1.2,-1.2)(1.73,-1.73)
        \psline(1.83,-1.83)(1.9,-1.9)
        \psline[ArrowInside=->](1.7,-1.7)(1.4,-1.4)\rput(1.6,-1.4){$a$}
        \psline[ArrowInside=->](2.1,-1.7)(2.4,-1.4)\rput(2.2,-1.4){$b$}
        \psline[ArrowInside=->](1.9,-2.5)(1.9,-1.9)\rput(1.98,-2.1){$c$}
                %%%
        \psline[border=2pt](2.6,-3.2)(1.9,-2.5)
        \psline(1.2,-3.2)(1.73,-2.67)
        \psline(1.83,-2.57)(1.9,-2.5)
        \psline[ArrowInside=->](1.4,-3)(1.7,-2.7)\rput(1.6,-3){$a'$}
        \psline[ArrowInside=->](2.4,-3)(2.1,-2.7)\rput(2.15,-3){$b'$}
    \end{pspicture}
 &\defineas \sum_{e} \Big[ F^{ab}_{a'b'}\Big]_{ce}
 \begin{pspicture}[shift=-1.05](1.1,-3.3)(2.5,-1.2)
        \scriptsize
        %Ellipse
        \psellipse[linecolor=black,border=0](1.8,-2.2)(0.2,.6) \rput(1.8,-1.5){$\omega_0$}
        \psline{>-}(1.61,-2.25)(1.61,-2.15)
        \psline[border=1.5pt](1.7,-2.1)(2.3,-2.4)
        \psline[ArrowInside=->](2.3,-2.4)(1.9,-2.2)
        \psline(1.55,-2.025)(1.3,-1.9)
        \psline[ArrowInside=->](1.3,-1.8)(1.3,-1.5)
        \psline[ArrowInside=->](1.3,-3)(1.3,-2.5)
        \psline(1.3,-3)(1.3,-1.5)
        \psline[ArrowInside=->](2.3,-1.8)(2.3,-1.5)
        \psline[ArrowInside=->](2.3,-3)(2.3,-2.5)
        \psline(2.3,-3)(2.3,-1.5)
        \rput(2.3,-1.3){$b$}
        \rput(1.3,-1.3){$a$}
        \rput(2.3,-3.2){$b'$}
        \rput(1.3,-3.2){$a'$}
        \rput(2.15,-2.125){$e$}
 \end{pspicture} \\
 &= \sqrt{\frac{d_c}{d_a d_b}} \delta_{a,a'}\delta_{b,b'}
 \begin{pspicture}[shift=-1.05](1.1,-3.3)(2.5,-1.2)
        \scriptsize
        \psline[ArrowInside=->](1.3,-3)(1.3,-1.5)
        \psline[ArrowInside=->](1.8,-3)(1.8,-1.5)
        \rput(1.8,-1.2){$b$}
        \rput(1.3,-1.2){$a$}
 \end{pspicture}
.
\end{align}
In other words, the definition (\ref{eq:Sace}) is constructed in a way to extract only the correlations associated with the anyonic charge lines that connect the two subsystems $A$ and $B$.
Hence, ACE is identically zero in the states of the form,
\begin{align}
\arho_{\text{sep}}\in V^{a_1,...,a_m}_{a_1,...,a_m}\otimes V^{b_1,...,b_n}_{b_1,...,b_n},
\end{align}
since $\arho_{\text{sep}}=D_{A:B}[\arho_{\text{sep}}]$. As we will see in Sec.\,\ref{sec:separability}, these are called separable states and their ALN vanishes as well.

\section{Anyonic partial transpose}
\label{sec:Anyonic partial transpose}

In this section, we introduce a diagrammatic implementation of the PT. 
As the reader may have noticed, in the diagrammatic formalism, we implicitly assume an arrow of time by representing anyon worldlines traveling forward in time as upward oriented lines. Hence, we reverse the arrow of time of one subsystem by exchanging the corresponding anyon lines vertically.

Let us begin with a diagrammatic implementation of the full transposition by exchanging the endpoints of the lines from top and bottom of a density matrix operator
\begin{align}
\label{eq:transpose}
\left[
\psscalebox{.85}{
 \pspicture[shift=-1.](-1,-1.1)(1,1.2)
  \footnotesize
%%%%% Box:
  \psframe[linewidth=0.9pt,linecolor=black,border=0](-0.8,-0.5)(0.8,0.5)
  \rput[bl]{0}(-0.1,-0.1){\normalsize $\arho$}
  \rput[bl]{0}(-0.22,0.7){$\mathbf{\ldots}$}
  \rput[bl]{0}(-0.22,-0.75){$\mathbf{\ldots}$}
%%%%% Line connections:
  \psset{linewidth=0.9pt,linecolor=black,arrowscale=1.5,arrowinset=0.15}
  \rput[bl]{0}(0.5,1.1){$a_n$}
  \psline(0.6,0.5)(0.6,1)
  \rput[bl]{0}(-0.7,1.1){$a_1$}
  \psline(-0.6,0.5)(-0.6,1)
  \rput[bl]{0}(0.5,-1.25){$a_n$}
  \psline(0.6,-0.5)(0.6,-1)
  \rput[bl]{0}(-0.7,-1.25){$a_1$}
  \psline(-0.6,-0.5)(-0.6,-1)
%%%%% Arrows:
  \psline{->}(0.6,0.5)(0.6,0.9)
  \psline{->}(-0.6,0.5)(-0.6,0.9)
  \psline{-<}(0.6,-0.5)(0.6,-0.9)
  \psline{-<}(-0.6,-0.5)(-0.6,-0.9)
\endpspicture
}\right]^T
\defineas
\psscalebox{.85}{
 \pspicture[shift=-1.2](-4,-1.4)(1,1.4)
  \footnotesize
%%%%% Box:
  \psframe[linewidth=0.9pt,linecolor=black](-0.8,-0.5)(0.8,0.5)
  \rput[bl]{0}(-0.1,-0.1){\normalsize $\arho$}
  \rput[bl]{0}(-0.22,0.7){$\mathbf{\ldots}$}
  \rput[bl]{0}(-0.22,-0.75){$\mathbf{\ldots}$}
%%%%% Line connections:
  \psset{linewidth=0.9pt,linecolor=black,arrowscale=1.5,arrowinset=0.15}
  \psline(0.6,0.5)(0.6,1)
  \psline(-0.6,0.5)(-0.6,1)
  \psline(0.6,-0.5)(0.6,-1)
  \psline(-0.6,-0.5)(-0.6,-1)
%%% transpose
  \psline(0.6,-1)(-1.8,-1.533)
  \psline(-1.8,-1.533)(-3.6,1)
  \psline(-0.6,-1)(-1.2,-1.133)
  \psline(-1.2,-1.133)(-2.72,1)
  \psline(0.6,1)(-1.8,1.533)
  \psline[border=2pt](-1.8,1.533)(-3.6,-1)
  \psline(-0.6,1)(-1.2,1.133)
  \psline[border=2pt](-1.2,1.133)(-2.72,-1)    
%%%%%% labels
  \rput[bl]{0}(-3.7,-1.3){$\bar a_n$}
  \rput[bl]{0}(-3.7,1.1){$\bar a_n$}
  \rput[bl]{0}(-2.8,-1.3){$\bar a_1$}
  \rput[bl]{0}(-2.8,1.1){$\bar a_1$}
%%%%% Arrows:
  \psline{->}(0.6,0.5)(0.6,0.9)
  \psline{->}(-0.6,0.5)(-0.6,0.9)
  \psline{-<}(0.6,-0.5)(0.6,-0.9)
  \psline{-<}(-0.6,-0.5)(-0.6,-0.9)
\endpspicture
}.
\end{align}
Given this observation, we propose a definition for PT. At the level of blocks, the anyonic PT looks like
\begin{align}
\label{eq:p_transpose}
\left[
\psscalebox{.85}{
 \pspicture[shift=-1](-1.1,-1.1)(1.1,1.1)
  \footnotesize
%%%%% Box:
  \psframe[linewidth=0pt,linecolor=white,border=0,fillstyle=solid,fillcolor=lightgray](-1,-0.5)(0,0.5)
    \psframe[linewidth=0.9pt,linecolor=black,border=0](-1,-0.5)(1,0.5)
    \rput[bl]{0}(-0.1,-0.1){\normalsize $\arho$}
  \rput[bl]{0}(-0.65,0.7){$\mathbf{\ldots}$}
  \rput[bl]{0}(0.25,0.7){$\mathbf{\ldots}$}
  \rput[bl]{0}(-0.65,-0.75){$\mathbf{\ldots}$}
  \rput[bl]{0}(0.25,-0.75){$\mathbf{\ldots}$}
%%%%% Line connections:
  \psset{linewidth=0.9pt,linecolor=black,arrowscale=1.5,arrowinset=0.15}
  \psline(0.1,0.5)(0.1,1)
  \psline(0.8,0.5)(0.8,1)
  \psline(-0.8,0.5)(-0.8,1)
  \psline(-0.1,0.5)(-0.1,1)
  \psline(0.1,-0.5)(0.1,-1)
  \psline(0.8,-0.5)(0.8,-1)
  \psline(-0.8,-0.5)(-0.8,-1)
  \psline(-0.1,-0.5)(-0.1,-1)
%%%%% labels:
  \rput[bl]{0}(-0.9,1.1){$a_1$}
  \rput[bl]{0}(-0.9,-1.3){$a_1$}
  \rput[bl]{0}(-0.35,1.1){$a_n$}
  \rput[bl]{0}(-0.35,-1.3){$a_n$}
  \rput[bl]{0}(0.08,1.1){$b_1$}
  \rput[bl]{0}(0.1,-1.3){$b_1$}
  \rput[bl]{0}(0.7,1.1){$b_n$}
  \rput[bl]{0}(0.7,-1.3){$b_n$}
%%%%% Arrows:
  \psline{->}(0.1,0.5)(0.1,0.9)
  \psline{->}(0.8,0.5)(0.8,0.9)
  \psline{->}(-0.8,0.5)(-0.8,0.9)
  \psline{->}(-0.1,0.5)(-0.1,0.9)
  \psline{-<}(0.1,-0.5)(0.1,-0.9)
  \psline{-<}(0.8,-0.5)(0.8,-0.9)
  \psline{-<}(-0.1,-0.5)(-0.1,-0.9)
  \psline{-<}(-0.8,-0.5)(-0.8,-0.9)
\endpspicture
}\right]^{T_A}
\defineas
\psscalebox{.85}{
 \pspicture[shift=-1.4](-3.2,-1.5)(1,1.5)
  \footnotesize
%%%%% Box:
  \psframe[linewidth=0pt,linecolor=white,border=0,fillstyle=solid,fillcolor=lightgray](-1,-0.5)(0,0.5)
    \psframe[linewidth=0.9pt,linecolor=black,border=0](-1,-0.5)(1,0.5)
  \rput[bl]{0}(-0.1,-0.1){\normalsize $\arho$}
  \rput[bl]{0}(-0.65,0.7){$\mathbf{\ldots}$}
  \rput[bl]{0}(0.25,0.7){$\mathbf{\ldots}$}
  \rput[bl]{0}(-0.65,-0.75){$\mathbf{\ldots}$}
  \rput[bl]{0}(0.25,-0.75){$\mathbf{\ldots}$}
%%%%% Line connections:
  \psset{linewidth=0.9pt,linecolor=black,arrowscale=1.5,arrowinset=0.15}
  \psline(0.1,0.5)(0.1,1)
  \psline(0.8,0.5)(0.8,1)
  \psline(-0.8,0.5)(-0.8,1)
  \psline(-0.1,0.5)(-0.1,1)
  \psline(0.1,-0.5)(0.1,-1)
  \psline(0.8,-0.5)(0.8,-1)
  \psline(-0.8,-0.5)(-0.8,-1)
  \psline(-0.1,-0.5)(-0.1,-1)
%%%% transpose
  \psline(-0.8,-1)(-1.4,-1.16)
  \psline(-1.4,-1.16)(-2.5,1)
  \psline(-0.1,-1)(-1.8,-1.5)
  \psline(-1.8,-1.5)(-3.075,1)
  \psline(-0.8,1)(-1.4,1.16)
  \psline[border=2pt](-1.4,1.16)(-2.5,-1)
  \psline(-0.1,1)(-1.8,1.5)
  \psline[border=2pt](-1.8,1.5)(-3.075,-1)
%%%%% labels:
  \rput[bl]{0}(-3.2,1.1){$\bar a_n$}
  \rput[bl]{0}(-3.2,-1.3){$\bar a_n$}
  \rput[bl]{0}(-2.6,1.1){$\bar a_1$}
  \rput[bl]{0}(-2.6,-1.3){$\bar a_1$}
  \rput[bl]{0}(0.08,1.1){$b_1$}
  \rput[bl]{0}(0.1,-1.3){$b_1$}
  \rput[bl]{0}(0.7,1.1){$b_n$}
  \rput[bl]{0}(0.7,-1.3){$b_n$}
%%%%% Arrows:
  \psline{->}(0.1,0.5)(0.1,0.9)
  \psline{->}(0.8,0.5)(0.8,0.9)
  \psline{->}(-0.8,0.5)(-0.8,0.9)
  \psline{->}(-0.1,0.5)(-0.1,0.9)
  \psline{-<}(0.1,-0.5)(0.1,-0.9)
  \psline{-<}(0.8,-0.5)(0.8,-0.9)
  \psline{-<}(-0.1,-0.5)(-0.1,-0.9)
  \psline{-<}(-0.8,-0.5)(-0.8,-0.9)
\endpspicture
}
\, .
\end{align}
As we see, we get braiding exchange of anyon worldlines within subsystem $A$.  
We then define the anyonic logarithmic negativity (ALN) as
\begin{align} 
    \label{eq:anyonic_neg_def}
    {\cal E}(A:B) &= \ln \norm{\arho^{T_A} }_1,
\end{align}
in terms of singular values of $\arho^{T_A}$.
We should note that
the anyonic PT defined above is not necessarily unique as we can braid clockwise or counterclockwise anyons at different positions. However,
different choices of braiding do not change the singular values and lead to the same value for the ALN. Hence, we do not need to make a specific choice. This arbitrariness is reminiscent of basis dependence of the partial transpose in conventional systems. 

We further notice that the order of the charge lines (from left to right) is reversed as we perform the PT. This however is not an issue, since we always contract $\arho^{T_A}$ with itself (or its Hermitian conjugate).
As a sanity check, we note that the above construction manifestly obeys the identity $\aTr\arho^2 =\aTr ( \arho^{T_A} \arho^{T_A\dag})$ which is valid for any definition of PT.

Finally, we should add that the above definition of anyonic PT is not necessarily  trace preserving nor Hermitian. To resolve this, one can perform local unitary transformations to make it trace preserving. Nevertheless, we technically do not need $\arho^{T_A}$ to be Hermitian or trace preserving as far as calculating the ALN is concerned. This is because ALN only depends on the singular values of 
$\arho^{T_A}$ 
(i.e., square roots of the spectrum of the Hermitian operator $\arho^{T_A\dag}\arho^{T_A}$).

%{\color{blue}
We finish this section by a warm-up example on how ALN is computed for a special case of an anyon dimer which fuses into the identity channel, as shown below,
\begin{align}
\label{eq:warm-up}
    \arho_{a}&= 
    \frac{1}{d_a}
    \begin{pspicture}[shift=-0.6](-0.2,0)(1.2,1.5)
        \scriptsize
        \psline[ArrowInside=->](0,0)(0.5,0.5)\rput(0,0.25){$a$}
        \psline[ArrowInside=->](1,0)(0.5,0.5)\rput(1,0.25){$\bar a$}
        \psline[ArrowInside=->](0.5,1)(0,1.5)\rput(0,1.25){$a$}
        \psline[ArrowInside=->,](0.5,1)(1,1.5)\rput(1,1.25){$\bar a$}
    \end{pspicture}.
\end{align}
Following the steps described above and taking the anyonic PT, we get
\begin{align}
    \label{eq:ptrans-warm-up}
    \arho^{T_A}_{a} = 
    \frac{1}{d_a}
    \begin{pspicture}[shift=-0.6](-1.2,0)(0.,1.5)
        \scriptsize
        % braid
        \psline(0,0)(-1,1.5)
        \psline[border=2pt](-1,0)(0,1.5)
        \psline[ArrowInside=->](0,0)(-0.33,0.5)\rput(-1,1.2){$\bar a$}
        \psline[ArrowInside=->](-1,0)(-0.66,0.5)\rput(-1,0.3){$\bar a$}
    \end{pspicture}\ .
\end{align}
In order to compute the ALN, we need to calculate the norm of $\arho^{T_1}$ which involves computing $\sqrt{\arho^{T_1\dag}\arho^{T_1}}$. Defining the square root in the diagrammatic approach may look difficult. However, we observe in this case that 
\begin{align}
    \arho^{T_A}_{a} (\arho^{T_A}_{a})^\dag = 
    \frac{1}{d_a^2}
    \begin{pspicture}[shift=-0.6](-0.2,0)(1.,1.5)
        \scriptsize
        \psline[ArrowInside=->](0,0)(0,1.5)\rput(0.2,0.75){$\bar a$}
        \psline[ArrowInside=->](0.7,0)(0.7,1.5)\rput(0.9,0.75){$\bar a$}
    \end{pspicture},
\end{align}
which is proportional to the identity operator. Therefore, we find the ALN to be
\begin{align}
    {\cal E}= \ln d_a,
\end{align}
which is non-zero when $a$ is non-Abelian. In view of the fact that $d_a$ can be thought of as the local \emph{anyonic} Hilbert space dimension, the state (\ref{eq:warm-up}) can then be regarded as anyonic version of maximally entangled state of two anyons. In the next section, we investigate general properties of ALN defined earlier. Subsequently, we will discuss more examples in Sec.~\ref{sec:Examples}.
%}

\section{General properties}
\label{sec:properties}

In order for an entanglement measure to be useful, it should satisfy several requirements~\cite{Vedral1997,Brus2002,Eisert_thesis}.  In this section,  we check these conditions for the ALN. We briefly list them here:
\begin{enumerate}[label=\alph*.]
    \item Vanishing for separable states;
    \item Invariance under local unitaries;
    \item Additivity;
    \item Monotonicity under LOCC;
    \item Computability, and continuity.
\end{enumerate}
We should note that one may impose other conditions as requirements for a useful entanglement measure. Here, we considered the most common ones from the quantum information literature.
Furthermore, the last condition is only necessary for practical purposes. For instance, a given quantity can be a measure of entanglement by satisfying the first four conditions, but it may be very difficult to compute it. In this regard, such a measure is not practically useful for quantifying  the entanglement.

\subsection{Zero entanglement for anyonic separable states}
\label{sec:separability}

Here, we need to show that PT acting on a separable state, defined in Eq.~(\ref{eq:separable}), does not change its norm (which is one). 
For concreteness, we consider a general form of a separable state in a tensor product Hilbert space $V^{a_1\cdots a_n}_{a_1\cdots a_n}\otimes V^{b_1\cdots b_m}_{b_1\cdots b_m}$,
\begin{align}
    \arho_{\text{sep}}=\sum_{\vec{e},\vec{e}',f,f'} p_{\vec{e},\vec{e}',ff'}\, \arho_A^{(\vec{e},f)} \otimes \arho_B^{(\vec{e}',f')},
\end{align}
where $\vec{e}$ and $\vec{e}'$ label the fusion channels in the fusion tree and $f$ and $f'$ denote the net fusion channels.
We note that the action of PT becomes a full transpose on one subsystem, i.e.,
\begin{align}
    \arho_{\text{sep}}^{T_A}={\sum_{\vec{e},\vec{e}',f,f'} p_{\vec{e},\vec{e}',ff'}\, (\arho_A^{(\vec{e},f)})^T \otimes \arho_B^{(\vec{e}',f')}}.
\end{align}
We observe that 
\begin{align}
    \label{eq:sep2}
    \arho_{\text{sep}}^{T_A} \arho_{\text{sep}}^{T_A\dag}
    &=\sum_{\vec{e},\vec{e}',f,f'} p_{\vec{e},\vec{e}',ff'}^2\, (\arho_A^{(\vec{e},f)})^T (\arho_A^{(\vec{e},f)})^{T\dag} \otimes (\arho_B^{(\vec{e}',f')})^2 \nonumber \\
    &=\sum_{\vec{e},\vec{e}',f,f'} p_{\vec{e},\vec{e}',ff'}^2\, (\bar \arho_A^{(\vec{\bar e},\bar f)})^2 \otimes (\arho_B^{(\vec{e}',f')})^2,
\end{align}
where $\bar{\cdots}$ on $\arho_A$ means the operator is turned upside-down, and as usual, $\bar{\cdots}$ over an anyon symbol denotes the corresponding anti-particle.
The first identity in the above follows from the orthogonality of anyonic states and the second identity follows from the property that 
\begin{align}
    \label{eq:rT_theta}
    (\arho_A^{(\vec{e},f)})^T = \theta_f^\ast\, \bar \arho_A^{(\vec{\bar e},\bar f)},
\end{align}
which diagrammatically means
\begin{align}
\psscalebox{0.8}{
\begin{pspicture}[shift=-2.1](-2.5,-3.5)(2,1)
\scriptsize
        %Top charges
        \psline[ArrowInside=->](.3,-.3)(0,-0.05)\rput(0,.2){$a_1$}
        \psline[ArrowInside=->](.3,-.3)(.6,0)\rput(.6,.2){$a_2$}
        \psline(.4,-.4)(.3,-.3)
        \psline[linestyle=dotted](.7,-.7)(.4,-.4)
        \rput(.6,-.4){$e_2$}
        \psline(.8,-.8)(.7,-.7)
        \psline(.8,-.8)(1.6,0)\psline[ArrowInside=->](1.3,-.3)(1.6,0)\rput(1.6,.2){$a_n$}
        \psline[ArrowInside=->](.8,-1.5)(.8,-.8)\rput(.6,-1.05){$f$}
        % transpose
        \psline(1.6,0)(0,1.1)
        \psline(0.6,0)(0,0.5)
        \psline(0,-0.05)(-1.8,-2.3)
        \psline(0,0.5)(-2.2,-2.3)
        \psline(0,1.1)(-2.7,-2.3)
        %lower piece
        %%%
        \rput(0,2.1){
        \psline[border=2pt](0,-5.5)(-2.7,-2.1)
        \psline[border=2pt](-0.0,-4.9)(-2.2,-2.1)
        \psline[border=2pt](0,-4.35)(-1.8,-2.1)
        \psline[ArrowInside=->](0,-4.35)(.3,-4.1)\rput(0,-4.6){$a_1$}
        \psline[ArrowInside=->](.6,-4.4)(.3,-4.1)\rput(.6,-4.6){$a_2$}
        \psline(.4,-4)(.3,-4.1)
        \psline[linestyle=dotted](.7,-3.7)(.4,-4)
        \psline(.8,-3.6)(.7,-3.7)
        \psline(.8,-3.6)(1.6,-4.4)\psline[ArrowInside=->](1.6,-4.4)(1.3,-4.1)\rput(1.6,-4.6){$a_n$}
        % transpose
        \psline(1.6,-4.4)(0,-5.5)
        \psline(.6,-4.4)(0.0,-4.9)
        }
    \end{pspicture}}
    =\theta_{f}^\ast 
\psscalebox{0.8}{
    \begin{pspicture}[shift=-2](-.5,-3.5)(1.8,.3)
    \scriptsize
        %Top charges
        \psline[ArrowInside=->](.3,-.3)(0,0)\rput(0,.2){$\bar a_1$}
        \psline[ArrowInside=->](.3,-.3)(.6,0)\rput(.6,.2){$\bar a_2$}
        \psline(.4,-.4)(.3,-.3)
        \psline[linestyle=dotted](.7,-.7)(.4,-.4)
        \rput(.6,-.4){$\bar e_2$}
        \psline(.8,-.8)(.7,-.7)
        \psline(.8,-.8)(1.6,0)\psline[ArrowInside=->](1.3,-.3)(1.6,0)\rput(1.6,.2){$\bar a_n$}
        \psline[ArrowInside=->](.8,-2.1)(.8,-.8)\rput(.6,-1.45){$\bar f$}
                %%%
        \rput(0,1.5){\psline[ArrowInside=->](0,-4.4)(.3,-4.1)\rput(0,-4.6){$\bar a_1$}
        \psline[ArrowInside=->](.6,-4.4)(.3,-4.1)\rput(.6,-4.6){$\bar a_2$}
        \psline(.4,-4)(.3,-4.1)
        \psline[linestyle=dotted](.7,-3.7)(.4,-4)
        \psline(.8,-3.6)(.7,-3.7)
        \psline(.8,-3.6)(1.6,-4.4)\psline[ArrowInside=->](1.6,-4.4)(1.3,-4.1)\rput(1.6,-4.6){$\bar a_n$}
        }
    \end{pspicture}}.
\end{align}
The above relation in turn yields
\begin{align}
    \norm{(\arho_A^{(\vec{e},f)})^T}_1 
    = \norm{\arho_A^{(\vec{e},f)}}_1 =1.
\end{align}
Finally, Eq.~(\ref{eq:sep2}) implies the norm invariance under PT and hence a zero ALN,
\begin{align}
    {\cal E}(\arho_{\text{sep}}) &= \ln \norm{\arho^{T_A}_{\text{sep}}}_1 
    =\ln \norm{\arho_{\text{sep}}}_1
    = 0.
\end{align}

%{\color{blue}
Regardless of separability, because of the triangle inequality for the one-norm we can put forward the following theorem.

\begin{thm}
The subspace of states with vanishing ALN forms a convex set.
\label{thm:convexity}
\end{thm}

\noindent\emph{Proof}:

 Suppose two density matrices $\arho_1$ and $\arho_2$ have vanishing ALN, i.e., $\norm{\arho_1^{T_A}}_1=\norm{\arho_2^{T_A}}_1=1$. Then, the ALN of any linear combination of these two states is also zero, since
\begin{align}
    1\leq \norm{\left(p\arho_1+(1-p)\arho_2\right)^{T_A}}_1 
    &= \norm{p\arho_1^{T_A}+(1-p)\arho_2^{T_A}}_1 \nonumber
    \\
    & \leq p \norm{\arho_1^{T_A}}_1 + (1-p) \norm{\arho_2^{T_A}}_1  \nonumber \\
    &=1,
\end{align}
where $0\leq p \leq 1$. Therefore, ${\cal E}\left(p\arho_1+(1-p)\arho_2\right)=0$.
%}

\hfill $\blacksquare$

\subsection{Invariance under local unitary transformation}
\label{sec:Unitary Invariance}

 Applying a local unitary operator must not change the entanglement measure.
 A local unitary is represented by 
 \begin{align}
 \arho \to (U_A\otimes U_B) \arho (U_A^\dag \otimes  U_B^\dag),
 \end{align}
where $U_s, \ s=A,B$ are unitary operators acting on subsystems $A$ and $B$, respectively. We are to show that
\begin{align} \label{eq:mono2}
\norm{ \left( (U_A\otimes U_B) \arho (U_A^\dag \otimes  U_B^\dag)\right)^{T_A} }_1 = \norm{\arho^{T_A}}_1.
\end{align}
We first note that
\begin{align}
\label{eq:unitary}
\left( (U_A\otimes U_B) \arho (V_A \otimes  V_B)\right)^{T_A}= 
(\bar V_A \otimes U_B) \arho^{T_A} (\bar U_A \otimes V_B),
\end{align}
where $\bar V_A$ ($\bar U_A$) denotes modified unitary operators which act from left (right) as opposed to their original form which acts from right (left). The above identity is a consequence of the pivotal property of unitary categories~\cite{Kitaev2006} which can be shown diagrammatically as,
\begin{align}
    \psscalebox{.8}{
     \begin{pspicture}[shift=-2](-3.2,-2.2)(1.1,2.2)
      \small
    %%%%% Box:
      \psframe[linewidth=0.9pt,linecolor=black,border=0](-1,-0.5)(1,0.5)
      \rput[bl]{0}(-0.15,-0.1){$\rho$}
      \rput[bl]{0}(-0.65,0.7){$\mathbf{\ldots}$}
      \rput[bl]{0}(0.25,0.7){$\mathbf{\ldots}$}
      \rput[bl]{0}(-0.65,-0.75){$\mathbf{\ldots}$}
      \rput[bl]{0}(0.25,-0.75){$\mathbf{\ldots}$}
    %%%%% Line connections:
      \psset{linewidth=0.9pt,linecolor=black,arrowscale=1.5,arrowinset=0.15}
      \psline(0.1,0.5)(0.1,1)
      \psline(0.1,0.5)(0.1,1)\psline(0.1,1.5)(0.1,1.9)
      \psline{->}(0.1,1.6)(0.1,1.85)
      \psline(0.8,0.5)(0.8,1)
      \psline(0.8,0.5)(0.8,1)\psline(0.8,1.5)(0.8,1.9)
      \psline{->}(0.8,1.6)(0.8,1.85)
      \psline(-0.8,0.5)(-0.8,1)
      \psline(-0.8,0.5)(-0.8,1)\psline(-0.8,1.5)(-0.8,1.9)
      \psline{->}(-0.8,1.6)(-0.8,1.85)
      \psline(-0.1,0.5)(-0.1,1)\psline(-0.1,1.5)(-0.1,1.9)
      \psline{->}(-0.1,1.6)(-0.1,1.85)
      \psline(0.1,-0.5)(0.1,-1)
      \psline(0.1,-1)(0.1,-0.5)\psline(0.1,-1.9)(0.1,-1.5)
      \psline{->}(0.1,-1.85)(0.1,-1.6)
      \psline(0.8,-0.5)(0.8,-1)
      \psline(0.8,-1)(0.8,-0.5)\psline(0.8,-1.9)(0.8,-1.5)
      \psline{->}(0.8,-1.85)(0.8,-1.6)
      \psline(-0.8,-0.5)(-0.8,-1)
      \psline(-0.8,-1)(-0.8,-0.5)\psline(-0.8,-1.9)(-0.8,-1.5)
      \psline{->}(-0.8,-1.85)(-0.8,-1.6)
      \psline(-0.1,-0.5)(-0.1,-1)
      \psline(-0.1,-1)(-0.1,-0.5)\psline(-0.1,-1.9)(-0.1,-1.5)
      \psline{->}(-0.1,-1.85)(-0.1,-1.6)
      %% unitary
      \psframe[linewidth=0.9pt,linecolor=black,border=0](-0.9,-1.5)(-0.02,-1)\rput[bl]{0}(-0.65,-1.4){$V_A$}
      \psframe[linewidth=0.9pt,linecolor=black,border=0](0.05,-1.5)(0.9,-1)\rput[bl]{0}(0.3,-1.4){$V_B$}
      \psframe[linewidth=0.9pt,linecolor=black,border=0](-0.02,1)(-0.9,1.5)\rput[bl]{0}(-0.65,1.1){$U_A$}
      \psframe[linewidth=0.9pt,linecolor=black,border=0](0.9,1)(0.05,1.5)\rput[bl]{0}(0.3,1.1){$U_B$}
      %%%% transpose
      \rput(0,-0.9){
      \psline(-0.8,-1)(-1.4,-1.16)
      \psline(-1.4,-1.16)(-3,2.2)
      \psline(-0.1,-1)(-1.8,-1.5)
      \psline(-1.8,-1.5)(-3.5,2.2)
      }
      \rput(0,0.9){
      \psline[border=2pt](-1.4,1.16)(-3,-2.2)
      \psline[border=2pt](-1.8,1.5)(-3.5,-2.2)
      \psline(-0.8,1)(-1.4,1.16)
      \psline(-0.1,1)(-1.8,1.5)
      }
    %%%%% Arrows:
      \psline{->}(0.1,0.5)(0.1,0.9)
      \psline{->}(0.8,0.5)(0.8,0.9)
      \psline{->}(-0.8,0.5)(-0.8,0.9)
      \psline{->}(-0.1,0.5)(-0.1,0.9)
      \psline{-<}(0.1,-0.5)(0.1,-0.9)
      \psline{-<}(0.8,-0.5)(0.8,-0.9)
      \psline{-<}(-0.1,-0.5)(-0.1,-0.9)
      \psline{-<}(-0.8,-0.5)(-0.8,-0.9)
    \end{pspicture}
}
=\psscalebox{.8}{
     \begin{pspicture}[shift=-2](-3.2,-2.2)(1.1,2.2)
      \small
    %%%%% Box:
      \psframe[linewidth=0.9pt,linecolor=black,border=0](-1,-0.5)(1,0.5)
      \rput[bl]{0}(-0.15,-0.1){$\rho$}
      \rput[bl]{0}(-0.65,0.7){$\mathbf{\ldots}$}
      \rput[bl]{0}(0.25,0.7){$\mathbf{\ldots}$}
      \rput[bl]{0}(-0.65,-0.75){$\mathbf{\ldots}$}
      \rput[bl]{0}(0.25,-0.75){$\mathbf{\ldots}$}
    %%%%% Line connections:
      \psset{linewidth=0.9pt,linecolor=black,arrowscale=1.5,arrowinset=0.15}
      \psline(0.1,0.5)(0.1,1)\psline(0.1,1.5)(0.1,1.9)
      \psline{->}(0.1,1.6)(0.1,1.85)
      \psline(0.8,0.5)(0.8,1)\psline(0.8,1.5)(0.8,1.9)
      \psline{->}(0.8,1.6)(0.8,1.85)
      \psline(-0.8,0.5)(-0.8,1)
      \psline(-0.1,0.5)(-0.1,1)
      \psline(0.1,-1)(0.1,-0.5)\psline(0.1,-1.9)(0.1,-1.5)
      \psline{->}(0.1,-1.85)(0.1,-1.6)
      \psline(0.8,-1)(0.8,-0.5)\psline(0.8,-1.9)(0.8,-1.5)
      \psline{->}(0.8,-1.85)(0.8,-1.6)
      \psline(-0.8,-0.5)(-0.8,-1)
      \psline(-0.1,-0.5)(-0.1,-1)
      %% unitary
      \psframe[linewidth=0.9pt,linecolor=black,border=0](0.05,-1.5)(0.9,-1)\rput[bl]{0}(0.3,-1.4){$V_B$}
      \psframe[linewidth=0.9pt,linecolor=black,border=0](0.9,1)(0.05,1.5)\rput[bl]{0}(0.3,1.1){$U_B$}     
      %%%% transpose
      \psline(-1.4,-1.16)(-2.5,1)\psline(-2.5,1)(-2.5,1.4)
      \psline(-2.5,1.9)(-2.5,2.3)\psline{-<}(-2.5,1.9)(-2.5,2.2)
      \psline(-1.8,-1.5)(-3.075,1)\psline(-3.075,1)(-3.075,1.4)
      \psline(-3.075,1.9)(-3.075,2.3)\psline{-<}(-3.075,1.9)(-3.075,2.2)
      \psline(-0.8,-1)(-1.4,-1.16)
      \psline(-0.1,-1)(-1.8,-1.5)
      \psline[border=2pt](-1.4,1.16)(-2.5,-1)\psline(-2.5,-1)(-2.5,-1.4)
      \psline(-2.5,-1.9)(-2.5,-2.3)\psline{->}(-2.5,-1.9)(-2.5,-2.2)
      \psline[border=2pt](-1.8,1.5)(-3.075,-1)\psline(-3.075,-1)(-3.075,-1.4)
      \psline(-3.075,-1.9)(-3.075,-2.3)\psline{->}(-3.075,-1.9)(-3.075,-2.2)
      \psline(-0.8,1)(-1.4,1.16)
      \psline(-0.1,1)(-1.8,1.5)
    %% unitary
      \psframe[linewidth=0.9pt,linecolor=black,border=0](-3.15,1.4)(-2.4,1.9)\rput[bl]{180}(-2.6,1.8){$V_A$}
      \psframe[linewidth=0.9pt,linecolor=black,border=0](-3.15,-1.4)(-2.4,-1.9)\rput[bl]{180}(-2.6,-1.5){$U_A$}
    %%%%% Arrows:
      \psline{->}(0.1,0.5)(0.1,0.9)
      \psline{->}(0.8,0.5)(0.8,0.9)
      \psline{->}(-0.8,0.5)(-0.8,0.9)
      \psline{->}(-0.1,0.5)(-0.1,0.9)
      \psline{-<}(0.1,-0.5)(0.1,-0.9)
      \psline{-<}(0.8,-0.5)(0.8,-0.9)
      \psline{-<}(-0.1,-0.5)(-0.1,-0.9)
      \psline{-<}(-0.8,-0.5)(-0.8,-0.9)
    \end{pspicture}\ ,
}
\label{eq:unitary_diagram}
\end{align}
where we identify the upside down operators with $\bar V_A$ and $\bar U_A$ in Eq.~(\ref{eq:unitary}). The crucial point is that $\bar V_A$ and $\bar U_A$ are also unitary operators. Finally, the fact that one-norm is unitary invariant then leads to Eq.~(\ref{eq:mono2}).

\subsection{Additivity}

  Entanglement of a composite system is equal to the sum of the entanglements of the constituting systems.
 To be more specific, let us consider two sets of anyons $V^{a_1\cdots a_n b_1 \cdots b_m}_{a_1\cdots a_n b_1 \cdots b_m}$ and $V^{a_1'\cdots a_p' b_1' \cdots b_q'}_{a_1'\cdots a_p' b_1' \cdots b_q'}$. A tensor product state in the combined space can be written as
 \begin{align}
 \arho=\arho_{AB} \otimes \arho_{A'B'},
 \end{align}
 where $\arho_{AB}$ and $\arho_{A'B'}$ describe entangled states in $V$ and $V'$, respectively.
This situation, for example, is realized by stacking two systems where $V^{a_1\cdots a_n}_{a_1\cdots a_n}$ and $V^{a_1'\cdots a_p'}_{a_1'\cdots a_p'}$ spaces belong to subsystem $A$ and similarly for the subsystem $B$. The additivity condition requires that
  \begin{align} \label{eq:additivity}
  {\cal E} (\arho_{AB} \otimes \arho_{A'B'}) = {\cal E} (\arho_{AB}) + {\cal E}(\arho_{A'B'}).
  \end{align}
    This condition is satisfied in the diagrammatic approach by definition, since the anyonic partial transpose involves manipulating each diagram separately. In other words, 
    \begin{align}
        \label{eq:ptrans_tensor}
        (\arho_{AB} \otimes \arho_{A'B'})^{T_{A\cup A'}}=
        \arho_{AB}^{T_A} \otimes \arho_{A'B'}^{T_{A'}},
    \end{align}
    i.e., 
    the partial transpose and tensor product commute.

\subsection{Monotonicity under LOCC}
\label{sec:monotonicity}

Here, we show that ALN is an entanglement monotone under the action of
local quantum operations and classical communication (LOCC).
LOCC generally refers to a multi-party process which consists of a sequence of steps where one party performs local measurements and communicates the result to other parties. As a result, the density matrix is mapped into
\begin{align}
    \rho \to \sum_i p_i \rho_i,
\end{align}
where $\rho_i$ denote a set of post measurement density matrices and  $p_i$'s are the probabilities associated with each outcome such that $\sum_i p_i=1$. The monotonicity condition states that an entanglement quantity $f$ must not increase on average over the set $\{\rho_i \}$; i.e.,
\begin{align}
    \label{eq:monotonicity}
    f(\rho) \geq \sum_i p_i f(\rho_i).
\end{align}
Instead of working directly with LOCCs and proving the above inequality, we make use of the theorem in Refs.~\cite{Vidal2000,Horodecki2005,Plenio2005} which implies that a convex function\footnote{Note that ALN is defined in terms of one-norm, hence, it is a convex function by definition.}
$f$ is LOCC monotone if and only if $f$ is
\begin{enumerate}[label=(\arabic*),itemsep=-1ex]
    \item invariant under local unitary operations,
    
    \item invariant under adding local ancilla in an arbitrary state at either subsystems,
    
    \item monotone under local projective measurements,
    
    \item affine 
    on mixtures of states
possessing local orthogonal (ancilla) flags, i.e., the function is equal to
its average as in
    \begin{align}
    \label{eq:affinity}
        f(\sum_i p_i \rho_i \otimes \ket{i_R}\!\bra{i_R})
        = \sum_i p_i  f(\rho_i),
    \end{align}
    where $\ket{i_R} \in \Hi_R$ denotes a set of local orthogonal flags and can be combined to be part of either subsystems $A$ or $B$.
\end{enumerate}

Although the monotonicity conditions are commonly defined for desnity matrix operators in matrix notation, it is straightforward to promote them to the anyonic version and express them in terms of tilde operators. Condition (1) was already discussed in Sec.\,\ref{sec:Unitary Invariance}. In Appendix\,\ref{app:lemmas}, we show that ALN meets conditions (2) and (3). Furthermore, we observe that
\begin{align}
    \norm{ \left( \sum\nolimits_i p_i \arho_i \otimes \Pi_R^{(i)} \right)^{T_{A}}}_1
    &=\norm{  \sum\nolimits_i p_i \arho_i^{T_A} \otimes \Pi_R^{(i)} }_1 \nonumber \\
    &= \sum\nolimits_i p_i  \norm{\arho_i^{T_A} \otimes \Pi_R^{(i)} }_1 
    \nonumber \\
    &= \sum\nolimits_i p_i  \norm{\arho_i^{T_A} }_1, 
\end{align}
which is the anyonic version of the affinity condition~(\ref{eq:affinity}) for the norm of PT.
Here, $\Pi_R^{(i)}$ refers to an orthonormal set of anyonic projection operators for ancilla and the system is partitioned as $A\cup (BR)$.
We note that the second identity follows from the orthogonality of projection operators, and the last line is a consequence of condition (2) (see also Appendix~\ref{app:lemmas} for more details). 

Having shown that anyonic PT satisfies conditions (1)-(4), we conclude that
\begin{align}
    \norm{ \arho^{T_A} }_1 \geq \sum_i p_i \norm{ \arho_i^{T_A} }_1,
\end{align}
which is the anyonic version of the monotonicity condition~(\ref{eq:monotonicity}).
Finally, we use the fact that logarithm is a concave function to arrive at
\begin{align}
    {\cal E}(\arho) \geq \sum_i p_i {\cal E}(\arho_i).
\end{align}

%%%%%%%%%%%%%%%%%%%%%%%%%%%%%%%%%%%%%%%%%%%%%%%%%%%%%%%%%%%
 \subsection{Computability and Continuity}
  An entanglement measure is useful in practice if it can be efficiently computed
for every state. 
For a generic anyonic density matrix represented diagrammatically, it is straightforward to perform the transformation rule (\ref{eq:p_transpose}), albeit it may lead to complicated diagrams.
Furthermore, the entanglement measure should be continuous.
There is not much to prove here, since PT is a linear operation and ALN is defined in terms of a one-norm which is algebraically a continuous function.
 
 %%%%%%%%%%%%%%%%%%%%%%%%%%%%%%%%%%%%%%%%%%%%%%

\section{Application to a single pair of anyons}
\label{sec:Examples}

To illustrate how our proposed anyonic PT can be carried out, we use the entangled state $\arho_{ab}$ that describes a pair of anyons $a$ and $b$ which fuse to the total charge $f$ with matrix coefficients $[p^f]_{\mu\mu'}$,
\begin{align}
\label{eq:ab_density}
    \arho_{ab}&= \sum_{f,\mu,\mu'} \frac{[p^f]_{\mu\mu'}}{d_f}\ket{a,b,\mu;f}\bra{a,b,\mu';f} \nonumber \\
    &= \sum_{f,\mu,\mu'} \frac{[p^f]_{\mu\mu'}}{\sqrt{d_a d_b d_f}}
    \begin{pspicture}[shift=-0.6](-0.2,0)(1.2,1.5)
        \scriptsize
        \psline[ArrowInside=->](0,0)(0.5,0.5)\rput(0,0.25){$a$}
        \psline[ArrowInside=->](1,0)(0.5,0.5)\rput(1,0.25){$b$}
        \rput(0.75,0.57){$\mu'$}
        \psline[ArrowInside=->](0.5,0.5)(0.5,1)\rput(0.35,0.75){$f$}
        \rput(0.7,1){$\mu$}
        \psline[ArrowInside=->](0.5,1)(0,1.5)\rput(0,1.25){$a$}
        \psline[ArrowInside=->,](0.5,1)(1,1.5)\rput(1,1.25){$b$}
    \end{pspicture},
\end{align}
where $\mu,\mu'=1,\cdots,N_{ab}^c$ denote multiplicity indices. This is a generalization of the example given at the end of Sec.~\ref{sec:Anyonic partial transpose}. We further assume that the above density matrix is normalized $\aTr(\arho)=\sum_{f} \Tr [p^f] =1$ and positive semi-definite which further constrains the matrix $[p^f]_{\mu\mu'}$ of coefficients. For instance, $\det [p^f]>0$ for every $f$ is a necessary condition. 
%In the rest of our paper, we omit fusion multiplicities $\mu$ in most cases for simplicity, unless stated otherwise.

We now apply the definition (\ref{eq:p_transpose}) to the anyonic dimer state $\arho_{ab}$ in Eq.~(\ref{eq:ab_density}), which gives
\begin{align}
    \label{eq:ptrans_def}
    \arho^{T_A}_{ab} = \sum_{f,\mu,\mu'} \frac{[p^f]_{\mu\mu'}}{\sqrt{d_a d_b d_f}}
    \begin{pspicture}[shift=-0.6](-1.2,0)(1.2,1.5)
        \scriptsize
        % braid
        \psline(0,0)(-1,1.5)
        \psline[border=2pt](-1,0)(0,1.5)
        \psline[ArrowInside=->](0,0)(-0.33,0.5)\rput(-1,1.2){$\bar a$}
        \psline[ArrowInside=->](-1,0)(-0.66,0.5)\rput(-1,0.3){$\bar a$}
        \psline[ArrowInside=->](0,0)(0.5,0.5)\rput(0.4,0.2){$a$}
        \psline[ArrowInside=->](1,0)(0.5,0.5)\rput(1,0.25){$b$}
        \rput(0.7,0.6){$\mu'$}
        \psline[ArrowInside=->](0.5,0.5)(0.5,1)\rput(0.35,0.75){$f$}
        \rput(0.7,0.95){$\mu$}
        \psline[ArrowInside=->](0.5,1)(0,1.5)\rput(0.35,1.35){$a$}
        \psline[ArrowInside=->,](0.5,1)(1,1.5)\rput(1,1.25){$b$}
    \end{pspicture}\ ,
\end{align}
that can further be simplified as
\begin{align}
    \arho^{T_A}_{ab} &= \sum_{f,\mu,\mu'} \frac{[p^f]_{\mu\mu'}}{\sqrt{d_a d_b d_f}}
    \begin{pspicture}[shift=-1.3](-0.7,-0.7)(0.9,1.6)
    \scriptsize
    \psline(-0.3,0)(-0.3,1.6)
    \psline(0.7,0)(0.7,1.6)
    \psline{->}(0.7,1)(0.7,1.45)\rput(0.85,1.35){$b$}
    \rput(0.55,1.1){$\mu$}
    \psarc(0.69,0.92){0.19}{-70}{82}
    \psline(-0.3,0.45)(0.63,0.7)
    \psline{->}(-0.3,0.45)(0.25,0.6)\rput(0.15,0.8){$f$}
    \psline{->}(-0.3,0.4)(-0.3,1.2) \rput(-0.45,1.1){$\bar a$}
    \psline{->}(0.7,0)(0.7,0.3) \rput(0.85,0.15){$\bar a$}
    \psline{->}(-0.3,0.)(-0.3,0.3) \rput(-0.45,0.2){$b$}
    \rput(-0.45,0.5){$\mu'$}
    \psline(-0.3,0)(0.7,-0.5)
    \psline[border=2pt](0.7,0)(-0.3,-0.5)
    \end{pspicture} \\
        &=\frac{1}{d_b} \sum_{c,\nu,\nu'}  [M^c]_{\nu\nu'}
    \sqrt{\frac{d_c}{d_a d_b} }
    \begin{pspicture}[shift=-0.6](-0.2,0)(1.2,1.5)
        \scriptsize
        \psline[ArrowInside=->](0,0)(0.5,0.5)\rput(0,0.25){$\bar a$}
        \psline[ArrowInside=->](1,0)(0.5,0.5)\rput(1,0.25){$b$}
        \rput(0.72,0.55){$\nu'$}
        \psline[ArrowInside=->](0.5,0.5)(0.5,1)\rput(0.35,0.75){$c$}
        \rput(0.7,1){$\nu$}
        \psline[ArrowInside=->](0.5,1)(0,1.5)\rput(0,1.25){$\bar a$}
        \psline[ArrowInside=->](0.5,1)(1,1.5)\rput(1,1.25){$b$}
    \end{pspicture}\, , 
    \label{eq:pt_dimer}
\end{align}
where
\begin{align}
    \label{eq:Mmat}
    [M^c]_{\nu\nu'}=
    \sum_{\substack{f,\mu,\mu'\\\sigma,\sigma',\delta,\delta'}} & [p^f]_{\mu\mu'}
    [A^{ab}_f]_{\mu\delta}
    [A^{ab}_f]^\ast_{\mu'\delta'} [R^{f\bar a}_b]_{\delta\sigma}
  \nonumber \\
    &\times
        [F^{\bar a f \bar a}_{c}]^\ast_{(b,\sigma,\nu),(b,\delta',\sigma')} 
        [R^{\bar a b}_c]_{\sigma'\nu'}^\ast,
\end{align}
and $A$-symbols are unitary matrices associated with the $A$-moves (see the definition in Eqs.~\eqref{eq:Amove} and \eqref{eq:Asymbol} of Appendix~\ref{app:diagrams}).
Clearly, this process does not preserve the anyonic trace, because
\begin{align}
    \label{eq:tr_pt}
    \aTr\arho_{ab}^{T_A} = \sum_c \frac{d_c}{d_b} \Tr [M^c]=  \theta_a,
\end{align}
where the last identity comes from directly evaluating the anyonic trace of the diagram (\ref{eq:ptrans_def}). 
As mentioned earlier, we can define a trace-preserving anyonic PT by performing a local unitary to absorb $\theta_a$.
However, this step is not really necessary
when evaluating the ALN, since we need to compute the one-norm of $\arho^{T_A}$ which can in turn be written in terms of square root of the Hermitized operator $\arho^{T_A}_{ab} \arho^{T_A\dag}_{ab}$ where the $\theta_a$ factor cancels out. Hence, 
we obtain
\begin{align}
    \label{eq:ab_LN_mu}
    {\cal E}(A:B) 
    = \ln \norm{\arho^{T_A}_{ab}}_1
    =\ln \sum_c \frac{d_c}{d_b} \norm{M^c}_1.
%    =\ln \sum_c \frac{d_c}{d_b} \norm{\widetilde{M}^c}_1.
\end{align}
We should note that ALN is independent of which subsystem we apply the partial transpose to.
In Appendix~\ref{app:proof Ta=Tb}, we prove that $\norm{\arho_{ab}^{T_A}}_1=\norm{\arho_{ab}^{T_B}}_1$.
Moreover, Eq.~(\ref{eq:ab_LN_mu}) is always non-negative, because
\begin{align}
    \label{eq:nonneg_LN}
    \sum_c \frac{d_c}{d_b} \norm{M^c}_1 \geq \sum_c \frac{d_c}{d_b} \theta_a^\ast \Tr[M^c] =1
\end{align}
where we use Eq.~(\ref{eq:tr_pt}) in the last equality.

Let us now look at some special limits of the above expression.
When the fusion channel is multiplicity free, the ALN is simplified into
\begin{align}
\label{eq:ab_LN}
    {\cal E} =  \ln \sum_c \frac{d_c}{d_b} \left| \sum_f p_f R^{f\bar a}_b [F^{\bar a f \bar a}_{c}]^\ast_{b,b} \right|.
\end{align}
When there is only one Abelian fusion channel  (e.g.~the identity sector), $[p^f]_{\mu\mu'}= \delta_{ef} \delta_{\mu,1} \delta_{\mu',1}$, Eq.~(\ref{eq:ab_LN_mu}) reduces into
\begin{align} \begin{split}
    %\label{eq:one_fusion_nu}
    {\cal E} &= \ln \sum_c \frac{d_c}{d_b} \norm{ [F^{\bar a e \bar a}_{c}]_{(b,1,\nu),(b,1,\nu')}}_1 \\
    &= \ln \sum_c  \frac{d_c}{d_b} N_{\bar{a}b}^c = \ln d_a,
    \label{eq:LN_abelian_f}
\end{split} \end{align}
where we use the fact that $F$-symbols are unitary matrices and $\sum_c N_{ab}^c d_c = d_a d_b$.
From this observation, we also deduce that if $a$ or $b$ is Abelian, then the ALN is identically zero.
From now on, we omit the multiplicity indices for simplicity unless stated otherwise.

As a reference for comparison with (\ref{eq:ab_LN}), we note that ACE of the dimer state (\ref{eq:ab_density}) without multiplicity is given by
\begin{align}
    S_{\text{ace}} = \ln d_a + \ln d_b - \sum_f p_f \ln d_f + \sum_f p_f \ln p_f
\end{align}
which is simply the mutual information.
This quantity however overestimates the entanglement. For instance, when $a=b$ and there is only one Abelian fusion channel, we get 
\begin{align}
    S_{\text{ace}} =2\ln d_a
\end{align}
which is twice the logarithmic negativity in Eq.~(\ref{eq:LN_abelian_f}).

%{\color{blue}
Later in this section, we study several examples of multiplicity-free fusion categories and observe that ALN only vanishes at an isolated point or a line in a higher dimensional parameter space.
This observation inspires us to put forward the following theorem and its subsequent conjecture.

\begin{thm}
The subspace of vanishing ALN in multiplicity-free dimer states 
is at least one dimension lower than the entire space of dimer states.
\label{thm:multi-free}
\end{thm}

We should note that this statement is in stark contrast with ordinary spins (qubits) where zero LN (positive PT) states occupy a finite volume of the entire space of states~\cite{PhysRevA.73.022109,*PhysRevA.72.032304,Beigi2010,Ye2009,Zyczkowski1}.

\noindent\emph{Proof:}

We begin by noting that the space of dimer states is $(n-1)$-dimensional, parameterized by a vector $\vec{p}=(p_1,p_2, \cdots,p_{n}) \in \mathbb{R}^{n}$ with a unit trace constraint $\sum_{f=1}^{n} p_f =1$.
For multiplicity-free theories, PT can be viewed as a linear map from $\mathbb{R}^{n}$ to $\vec{m}=(m_1,\cdots,m_{n})\in \mathbb{C}^{n}$
where the complex valued components are defined in Eq.\,(\ref{eq:Mmat}), subject to the constraint $\sum_{c=1}^{n} m_c=1$~\footnote{Compared to (\ref{eq:Mmat}) we choose a normalization factor by including $\theta_a^\ast$ in the definition of $m_c$}. Alternatively, the linear relation between $\vec{m}$ and $\vec{p}$ can be recast as
\begin{align}
    m_i = 
    \sum_{j=1}^{n} \Delta_{ij} p_j,
\end{align}
where $\Delta$ is an $n\times n$ complex matrix which depends on $R$ and $F$ symbols as in Eq.\,(\ref{eq:Mmat}).
Because of the unit trace constraint on $\vec{m}$, the zero ALN condition $\sum_i |m_i|=1$ is met only when $m_i$'s are collinear on the real positive axis, i.e. $\text{Im}\ m_i=0$ for every $i$. In terms of $\Delta$, we can write these conditions as
\begin{align}
    \label{eq:vanishing_ALN}
    \sum_{j=1}^{n} \text{Im}[\Delta_{ij}] p_j = 0.
\end{align}
Therefore, the dimension of subspace of states with vanishing ALN depends on the rank of $\text{Im}[\Delta]$ and is given by
\begin{align}
\label{eq:r0}
r_0=n-1-\text{rank}(\text{Im}[\Delta]).
\end{align}
Notice that the unit trace requirement already imposes a constraint on $\text{Im}[\Delta]$ matrix, $\sum_{i,j=1}^{n} \text{Im}[\Delta_{ij}] p_j=0$, which makes it rank deficient. This in turn guarantees $r_0\geq 0$.
In general, $r_0$ is not \emph{a priory} known and depends on the dimer state and a given category. Nevertheless, it is unlikely that $\text{Im}[\Delta]$ will be a zero matrix; hence, $0\leq r_0 < n-1$, which is lower-dimensional than $(n-1)$-dimensional parameter space.

\hfill $\blacksquare$

In the case of $\text{rank}(\text{Im}[\Delta])=n-1$, there is only one solution to Eq.~(\ref{eq:vanishing_ALN}). Due to Theorem~\ref{thm:convexity}, the ALN is zero at a single point, and there is no choice but the separable state which is given by the probability coefficients in Eq.~(\ref{eq:thm}), written below.

\begin{corr}
When the subspace of vanishing ALN is zero-dimensional, ALN vanishes only at the separable state where
\begin{align}
    \label{eq:thm}
    p_f=\frac{d_f}{d_a d_b}.
\end{align}
In other words, zero ALN is a necessary and sufficient condition for separability in this case.
\end{corr}

We further believe that Theorem~\ref{thm:multi-free} can be generalized to categories with fusion multiplicities, although we do not present a rigorous proof here. 
% of the above conjecture here, we give a heuristic argument that the ALN of multiplicity-free categories vanishes  over a lower dimensional subspace (which are then of zero measure)

\begin{conjecture}
The subspace of vanishing ALN for dimer states forms a zero measure set.
\label{conj:multiplicity}
\end{conjecture}

%}

In the remainder of this section, we compute the ALN of the toy density matrix (\ref{eq:ab_density}) for some special cases in Ising anyons, Fibonacci category, $su(2)_k$, and $su(3)_k$ theories. As we will see, ALN vanishes along a line in the parameter space of dimer state of two spin-$1$ anyons in $su(2)_4$ and two spin-$8$ anyons in $su(3)_3$, otherwise, it equals zero at a single point.

\subsection{Ising anyons}

The Ising$^{(\nu)}$ anyon models~\cite{Kitaev2006,Nayak2008,BondersonGcrossed}
 contain three topological charges $\{I, \sigma, \psi\}$ with the following fusion rules
\begin{align}
	\psi\otimes\psi=I,
	\quad \sigma\otimes\psi = \psi\otimes\sigma=\sigma,
	\quad \sigma\otimes\sigma=I\oplus \psi.
\end{align}
Here, $\nu$ is an odd integer which labels the eight distinct Ising anyon models such that $\nu \sim \nu +16$, i.e., $\nu$ is defined modulo $16$. The quantum dimensions are
\begin{align}
    d_I= d_\psi=1, \quad d_\sigma=\sqrt{2}.
\end{align}
The nontrivial $F$-symbols are
\begin{align} \begin{aligned}
&
	F^{\psi\sigma\psi}_\sigma=F^{\sigma\psi\sigma}_\psi=-1, 
	\\
	&
	\left[F^{\sigma\sigma\sigma}_\sigma\right]_{ab}=
	\frac{\varkappa_{\sigma}}{\sqrt{2}} \left[
	\begin{matrix}
		1 & 1\\
		1 & -1
	\end{matrix}
\right],
	\label{eq:IsingF}
\end{aligned} \end{align}
where the latter matrix is in the $\{I, \psi \}$ basis, i.e., $a,b= I, \psi$, and $\varkappa_{\sigma}=(-1)^{\frac{\nu^2-1}{8}}$ is the Frobenius-Schur indicator of $\sigma$.
Furthermore, the $R$-symbols are
\begin{align} \begin{aligned}
	R^{\psi\sigma}_\sigma &=R^{\sigma\psi}_\sigma=(-i)^{\nu},\\
	R^{\sigma\sigma}_{I}&=\varkappa_{\sigma} e^{-i\frac{\pi}{8}\nu}, \quad R^{\sigma\sigma}_\psi=\varkappa_{\sigma} e^{i\frac{3\pi}{8}\nu}.
\end{aligned} \end{align}
The topological twist factor $\theta_\sigma=e^{i\frac{\pi }{8}\nu}$ uniquely distinguishes the eight distinct Ising$^{(\nu)}$ anyon models, as does the chiral central charge $c\bmod8 = \frac{\nu}{2}$.

\begin{figure}
\centering
    \includegraphics[scale=.7]{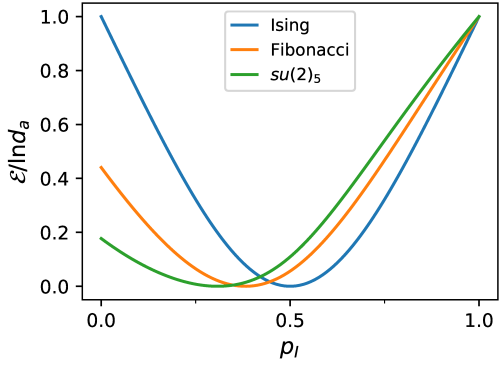}
    \caption{Anyonic logarithmic negativity (\ref{eq:ab_LN}) of two entangled anyons as a function of $p_I$, the probability of identity fusion charge in Eq.~(\ref{eq:ab_density}). Here, $a=b=\sigma$ for the Ising category, $a=b=\tau$ for the Fibonacci category, and $a=b=2$ for $su(2)_5$ theory. }
    \label{fig:IsingFibonacci}
\end{figure}

As our first example, we consider the dimer state of two Ising anyons, i.e., $a=b=\sigma$ in Eq.~(\ref{eq:ab_density}). Upon plugging in the anyon data for Eq.~(\ref{eq:ab_LN}), we arrive at
\begin{align}
    {\cal E} = \frac{1}{2} \ln [2 ( p_I^2 + p_\psi^2)],
\end{align}
where $p_I+ p_\psi=1$. 
As we see in this case, ${\cal E}$ 
%only vanishes 
vanishes only
when $p_I=p_\psi=1/2$,
 corresponding to Eq.~(\ref{eq:thm}). In other words, there is no other (in-)separable states with zero ALN. 

As our second example, we take $a=\sigma$ and $b=\psi$. In this case, the total fusion channel is fixed to be $f=\sigma$ and ${\cal E}=0$.

\begin{figure}
\centering
    \includegraphics[scale=.7]{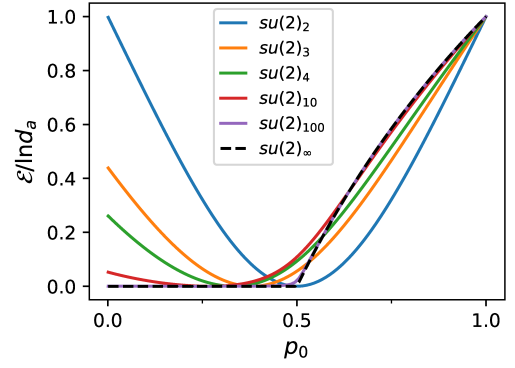}
    \caption{Anyonic logarithmic negativity (\ref{eq:ab_LN}) of two spin-1/2 anyons of $su(2)_k$ theory as a function of $p_0$, the probability of fusing into identity charge of $\rho_{ab}$ in Eq.~(\ref{eq:ab_density}).
    $su(2)_\infty$ corresponds to the LN of the Werner state of ordinary spin-$1/2$'s given in Eq.~(\ref{eq:LN-Werner}).
    Notice that ALN vanishes at a single point $p_0= d_{\frac{1}{2}}^{-2}$ for $su(2)_k$ anyons, whereas it vanishes over the range $p_{0}<1/2$  for the Werner state. }
    \label{fig:spinhalf}
\end{figure}

%%%%%%%%%%%%%%%%%%%%%%%%%%%%%%%%%%%%%%%%%%%%%%
\subsection{Fibonacci anyons}
The Fibonacci category has two topological charges $\{ I,\tau \}$, with the following non-trivial fusion rule
\begin{align}
 \tau \otimes \tau = I \oplus \tau.
\end{align}
The quantum dimensions are given by
\begin{equation}
d_I = 1, \quad d_\tau = \phi,
\end{equation}
where $\phi = \frac{1 + \sqrt{5}}{2}$ is the golden ratio and the non-trivial $F$-symbol and $R$-symbol are
\begin{align}
	\left[F^{\tau\tau\tau}_\tau\right]_{ab}=
	\begin{bmatrix}
	    \phi^{-1} & \phi^{-1/2} \\
	    \phi^{-1/2} & - \phi^{-1}
	\end{bmatrix},
\end{align}
where $a,b=I,\tau$,
and 
\begin{align}
	R^{\tau\tau}_I= e^{-i \frac{4 \pi}{5}},
	\quad
    R^{\tau\tau}_\tau= e^{i \frac{3 \pi}{5}},
\end{align}
respectively.
The twist factor of the Fibonacci anyon is given by $\theta_\tau = e^{i \frac{4 \pi}{5}}$.

Here, we consider the dimer state of two Fibonacci anyons, i.e., $a=b=\tau$ in Eq.~(\ref{eq:ab_density}) which fuses to $f=I,\tau$. Upon plugging in the anyon data for Eq.~(\ref{eq:ab_LN}), we get
\begin{align}
    {\cal E} =  \ln \frac{1}{\phi} \Big[
     \big| p_I + p_\tau R^{\tau\tau}_\tau \big| + \big| p_I \phi - p_\tau R^{\tau\tau}_\tau \big|
    \Big].
\end{align}
The two extreme limits are
when $p_I=1$,
\begin{align}
    {\cal E} = \ln \phi,
\end{align}
and
when $p_\tau=1$,
\begin{align}
    {\cal E} = \ln \frac{2}{\phi}.
\end{align}
Figure \ref{fig:IsingFibonacci} shows how ALN varies in the Ising and Fibonacci categories. Similar to the Ising anyons, ALN in the Fibonacci case vanishes only at a single point defined in Eq.~(\ref{eq:thm}).

\begin{figure*}
\centering
    \includegraphics[scale=1.2]{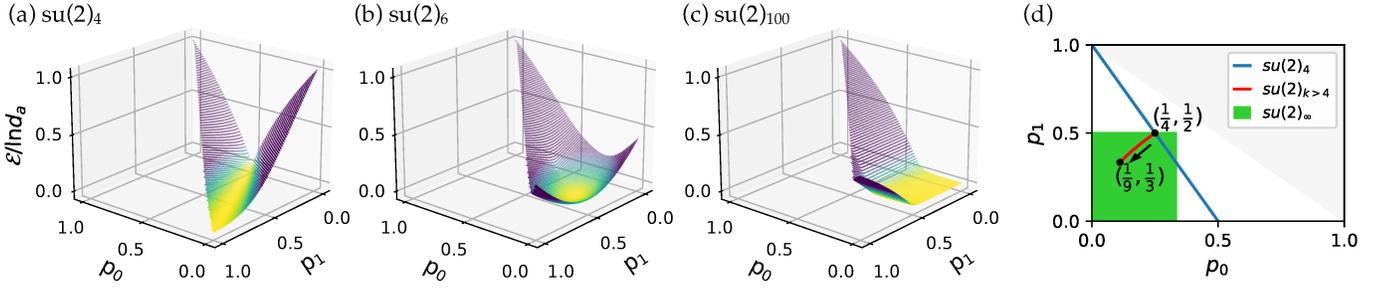}
    \caption{ 
    (a)-(c) Anyonic logarithmic negativity (\ref{eq:ab_LN}) of two spin-$1$ anyons (\ref{eq:ab_density}) for various $su(2)_k$ categories as a function of $p_0$ and $p_1$, the probability of fusing into spin-$0$ and spin-$1$ charges, respectively.
    (d) Zero locus of ALN for two spin-$1$ in all $su(2)_k$ theories.
    Notice that ALN equals zero along the blue line $p_1=1-2 p_0$ for $k=4$ theory, whereas it only vanishes at a single point for $k>4$ categories. The latter (isolated) zero moves along the red line as shown by the arrow in panel (d) as $k$ is swept from $k=4$ to $k\to \infty$. 
    In panel (d), $su(2)_\infty$ corresponds to the LN of two ordinary spin-$1$'s, which is identically zero in the green region $p_0\leq 1/3$, $p_1\leq 1/2$. Here, the gray shaded area is prohibited since $p_0+p_1\leq 1$. 
    }
    \label{fig:spin_one}
\end{figure*}
%%%%%%%%%%%%%%%%%%%%%%%%%%%%%%%%%%%%%%%%%%%%%%
\subsection{\texorpdfstring{$su(2)_k$}{su(2) lvl k} anyons}
\label{sec:examples-su2}

The deformed versions of the $su(2)$ spin models where the anyons are labeled by the first $k+1$ (generalized) angular momenta $\{0,\frac{1}{2},1,\frac{3}{2},\cdots,\frac{k}{2} \}$. They obey the fusion rule
\begin{align}
    j_1 \otimes j_2=
    \bigoplus_{j=|j_1-j_2|}^{\min\{j_1+j_2,k-j_1-j_2\}} j,
\end{align}
and their quantum dimensions are given by
\begin{align}
    d_j= \frac{\sin\frac{\pi(2j+1)}{k+2}} {\sin\frac{\pi}{k+2}}.
\end{align}
In the above notation, the identity sector is labeled by zero, i.e., $I\equiv 0$. The $F$ and $R$ symbols of this category are provided in Appendix~\ref{app:Fsymbols}.

Let us first consider an entangled state (\ref{eq:ab_density}) of two spin-2's in the $su(2)_5$ category associated with the following fusion rule,
\begin{align}
    2 \otimes 2 &= 0\oplus 1, 
\end{align}
which is shown as a green curve in Fig.~\ref{fig:IsingFibonacci}. By comparing the values of ALN at $p_0=0$ for different categories, we observe that the fusion channel with larger quantum dimension leads to a smaller ALN. 
Next, we consider a mixed state of two spin-$\tfrac{1}{2}$'s with the two fusion channels, 
\begin{align}
    \tfrac{1}{2} \otimes \tfrac{1}{2} = 0 \oplus 1.
\end{align}
Figure~\ref{fig:spinhalf} shows how the ALN for different $su(2)_k$ categories changes as the probability is tuned from the spin-1 channel towards the spin-0 (identity) channel. We observe that the entanglement in the spin-1 fusion channel decreases substantially as the level of the theory, $k$, is increased, while the negativity of the identity channel remains finite at $\ln d_{\frac{1}{2}}=\ln 2\cos(\pi/(k+2))$. Finally, as $k\gg 1$ is tuned towards the ordinary spins, it approaches the Werner state~\cite{Werner1989} of ordinary spin-$\tfrac{1}{2}$ (qubit) systems,
\begin{align}
    \rho_{w}=  p_0 \ket{s}\!\bra{s} + \left(\frac{1-p_0}{3}\right)\sum_{i=0,\pm} \ket{t_i}\!\bra{t_i}, 
\end{align}
where $\ket{t_0}= (\ket{\uparrow\downarrow}+\ket{\downarrow\uparrow})/\sqrt{2}$, $\ket{t_+}=\ket{\uparrow\uparrow}$, and $\ket{t_-}=\ket{\downarrow\downarrow}$ are spin triplet states and $\ket{s}= (\ket{\uparrow\downarrow}-\ket{\downarrow\uparrow})/\sqrt{2}$ denotes the spin singlet state.
The corresponding logarithmic negativity is given by
\begin{align}
    \label{eq:LN-Werner}
    {\cal E}(\rho_w)=\ln \left( \frac{1}{2}+ p_{0}+\left|\frac{1}{2}-p_0\right| \right),
\end{align}
which is shown as the dashed line in Fig.~\ref{fig:spinhalf}. 
We should note that all states with $p_0<1/2$ are separable for ordinary spins. This is in contrast with any theory of finite $k$ where ALN vanishes only at one point. 
The separable point of $su(2)_k$ is determined by (\ref{eq:thm}), that is
\begin{align}
    p_{0}^\ast= \frac{1}{[2\cos(\pi/(k+2))]^2},
\end{align}
which approaches $1/4$ in the $k\to \infty$ limit. This is clearly away from the sudden death point $p_0=1/2$ of the Werner state. Therefore, what happens to the ALN curves as we increase $k$ in $su(2)_k$ is that they get flatter and flatter near $p_0^\ast$, i.e., higher order derivatives $\frac{\partial^n {\cal E}}{\partial p_0^n}$ vanish at $p_0=p_0^\ast$. As shown in Fig.~\ref{fig:spinhalf}, $su(2)_{100}$ is already quite close to the Werner state.

%{\color{blue}
Let us now consider mixed states of two spin-$1$'s with three fusion channels
\begin{align}
    1\otimes 1 = 0 \oplus 1 \oplus 2,
\end{align}
for $k\geq 4$. This implies that the space of dimer states  is two dimensional, $(p_0,p_1)$.
 In particular, for $k=4$ we find that
\begin{align}
    {\cal E} = \ln \left[\frac{1}{2} \sum_{s=\pm 1} \left| p_0-p_2+ s e^{-i\frac{\pi}{3}} p_1 \right| + |p_0+p_2| \right],
\end{align}
where ALN is identically zero for $p_0=p_2=(1-p_1)/2$ (See Fig.~\ref{fig:spin_one}(a)). 
We observe two important differences from
the other cases we studied so far: First, the subspace of states with zero ALN is not simply an isolated point but a one-dimensional space. Second, this family of states is clearly not a product state or separable, while their ALN is zero.
To put it in the context of Theorem~\ref{thm:multi-free}, we find the $\Delta$ matrix to be
\begin{align}
    \Delta=\frac{1}{2} \begin{pmatrix}
    \theta & 1 & -\theta \\
    2 & 0 & 2 \\
    -\theta & 1 & \theta
    \end{pmatrix},
\end{align}
where $\theta=e^{i\frac{2\pi}{3}}$. We see that $\text{rank}(\text{Im}[\Delta])=1$, and hence, using Eq.~(\ref{eq:r0}), the dimension of the subspace of vanishing ALN is $r_0=1$.

We further numerically compute the ALN for $k>4$ and  typical results for level $k=6$ and $100$ are shown in Fig.~\ref{fig:spin_one}(b) and (c). We note that in these cases ALN vanishes only at one point as in Eq.~(\ref{eq:thm}).
To summarize the spin-$1$ results, we plot the zero locus of ALN in Fig.~\ref{fig:spin_one}(d). For reference, we also show the zero locus of two ordinary spin-$1$'s which covers a finite two-dimensional region $p_0\leq 1/3, p_1\leq 1/2$ (green region). Again, the crucial difference between anyonic states and ordinary spin states is that ALN only vanishes at points or along a line which are measure-zero sets in two-dimensional space.
Similar to the previous case of two spin-$\frac{1}{2}$ anyons, here also the ALN surface becomes flatter and flatter around the separable point (\ref{eq:thm}), as we increase $k$ (e.g., Fig.~\ref{fig:spin_one}(c)).
%}
% The existence of a single point where the ALN vanishes in the above examples, is again in favor of  our earlier conjecture.
% However, as we will see in an example below, our conjecture does not hold for dimer states with multiplicities.

\subsection{\texorpdfstring{$su(3)_3$}{su(3) lvl 3} anyons}
\label{sec:examples-su3}

We study a subset of the anyons within the $su(3)_3$ category as a simple theory with fusion multiplicity. The four anyons $\{1, 8, 10, \overline{10} \}$ are closed under fusion, and has quantum dimensions~\cite{Ardonne_2010},
\begin{align}
    d_1 &= d_{10} = d_{\overline{10}}= 1, \qquad
    d_8 = 3,
\end{align}
where the fusion rules are given by
\begin{align} \begin{aligned}
    8 \otimes 8 &= 1 \oplus 8 \oplus 8 \oplus 10 \oplus \overline{10}, \\
    8 \otimes 10 &= 8 \times \overline{10} = 8, \\
    10 \otimes 10 &= \overline{10}, \qquad
    \overline{10} \otimes \overline{10} = 10, \qquad
    \overline{10} \otimes 10 = 1.
\end{aligned} \end{align}
The crucial point here is that we have a non-trivial multiplicity $N_{88}^8=2$.
The $F$ and $R$ symbols are provided in Appendix~\ref{app:Fsymbols}.

We consider a state in the form of Eq.~(\ref{eq:ab_density}) where $a=b=8$ and we limit the fusion channel only to $f=8$ with two multiplicities. This density matrix is fully described by a $2 \times 2$ matrix 
\begin{align}
    [p^8] = 
    \begin{pmatrix}
    p^{8}_{11} & p^{8}_{12} \\
    p^{8}_{21} & p^{8}_{22}
    \end{pmatrix},
    \label{eq:su3}
\end{align}
in the Hilbert space $V^{88}_{88}$ and is characterized by three real parameters:
\begin{align}
    p^{8}_{11} &= p, \qquad
    p^{8}_{22} = 1-p, \nonumber \\
    p^{8}_{12} &=  p^{8\ast}_{21} = q_r + i q_i.
\end{align}
where $q_i, q_r\leq 1/2$ to ensure positive semi-definiteness of $\arho_{ab}$.
Plugging in the anyon data  for Eq.~(\ref{eq:ab_LN_mu}), we get
\begin{align}
    {\cal E}= \ln\left[ 1+
    \frac{|2p-1|}{3} +
    \frac{1}{6} \sum_{s=\pm 1} 
    \left|2p-1+2\sqrt{3} s q_r\right|
    \right].
    \label{eq:LN_su3}
\end{align}
(See Appendix~\ref{app:Fsymbols} for details.)
It is interesting to note that ALN does not depend on $q_i$ and vanishes when $p=1/2$, $q_r=0$ for arbitrary values of $q_i$. This gives not just a point but a family of states with vanishing ALN which lives on a line in the three-dimensional parameter space $(p,q_r,q_i)$. Similar to the case of two spin-$1$'s in $su(2)_4$, here also we get a family of states with zero ALN which is clearly not a product state or separable. We also note that this is consistent with Conjecture~\ref{conj:multiplicity}, since a one-dimensional space  occupies a measure zero volume in three-dimensional parameter space. 

\section{Conclusions}
\label{sec:Conclusions}

In summary,
we proposed a way to incorporate PT as an operation (braiding) on anyonic density matrices. We examined this construction in terms of standard requirements for faithful measures of entanglement including invariance under local unitaries and projection operators, monotonicity under LOCC, and additivity, and showed that it satisfies them all.
Moreover, ALN manifestly vanishes for separable anyonic states, which in the language of anyons corresponds to states where there is no non-trivial anyon line connecting the two parties. We then applied the anyonic PT to several examples of anyon dimers, where we found that although the PT depends on some 2D properties such as a choice of (over or under) braiding but the final result does not depend on these details. In other words, the notion of entanglement in anyons is algebraic beyond the dimensionality. The mentioned dependence is reminiscent of a similar phenomenon in the standard PT for qubits where the operation itself is basis dependent while the final quantities such as LN or topological invariants~\cite{Shap_unoriented} are basis independent. Looking at examples, we proved that in the case of multiplicity-free dimers, ALN vanishes only at lower-dimensional convex subspaces of the parameter space of density matrices. Specifically, when this subspace is zero-dimensional, this theorem implies that there is only one point with vanishing ALN which corresponds to a disconnected anyon diagram. We further argued that the assumption of multiplicity-free may not be  essential, and conjectured that our theorem can be generalized to categories with fusion multiplicities. For instance, in the case of $su(3)_3$ which is a category with a two-fold fusion multiplicity, we find that ALN vanishes along a line in a three-dimensional parameter space.

There are several new avenues for future research. ALN is known to provide an upper bound on distillable entanglement in qubit systems~\cite{Bennett1996a,Bennett1996b,Horodecki1997_distillation}. It would be interesting to figure out what are the analogs of distillation protocols for anyons and whether in this case ALN gives any bound on the amount of distillable entanglement. 
So far, we have studied the entanglement in braided tensor categories. Nevertheless, it is tempting to believe that our formalism can also be applied to entangled states of anyon symmetry defects such as the ones realized in G-crossed categories~\cite{BondersonGcrossed}, e.g., parafermion dimers in $\Z_N^{(p)}$ categories. We think that the anyonic PT may be defined in terms of the G-crossed braiding. Exploring such possibilities for generalizing the current formalism is worth pursuing.
Furthermore, our proposed diagrammatic implementation of PT may be adapted to implement other manipulations of anyonic density matrices such as realignment~\cite{realign1,realign2},
reflected entropy~\cite{Dutta2019,Zou2020},
and 
odd entropy 
\cite{Tamaoka}.
It would be interesting to find possible similarities and differences among these different measures.

Throughout this paper, we focus more on methodology, i.e., developing a framework to implement the PT in anyonic systems, rather than studying various physical phenomena in anyon models. 
A more comprehensive study on the application of this method to other many-anyon models such as anyon chains would be worth pursuing both from the standpoint of having more benchmark examples as well as uncovering entanglement structures within different strongly interacting many-body systems. 
In a sequel paper, we apply the anyonic PT to construct partition functions of anyon chains on non-orientable spacetime manifolds (i.e., an anyonic generalization of~\cite{Shap_unoriented}) and use the ALN to derive the topological entanglement negativity~\cite{Wen2016_1}.
Another future direction along this line may be the implementation of the anyonic PT in versatile numerical approaches such as the density matrix renormalization group (DMRG). For example, it would be interesting to investigate whether ALN in translationally symmetric anyonic chains (e.g., the golden chain~\cite{PhysRevLett.98.160409}) obeys the expected scaling form predicted by the conformal field theory~\cite{Calabrese2012}. 

\section*{Acknowledgments}
The authors would like to acknowledge insightful discussions with 
Parsa Bonderson, Meng Cheng, 
Paul Fendley, Christina Knapp, Ryohei Kobayashi, Max Metlitski, John Preskill, T.~Senthil, Ryan Thorngren, 
Ashvin Vishwanath, and Xueda Wen.
HS was supported by the Simons Collaboration on Ultra-Quantum Matter, which is a grant from the Simons Foundation (651440).
RM is supported by the National Science Foundation No.\ DMR-1848336.
SR is supported by a grant from the Simons Foundation (Award Number: 566116).
This works was initiated at
Aspen Center for Physics 2018 Winter Conference
``Field Theory Dualities and Strongly Correlated Matter"
March 18-24, 2018.

\appendix
%%%%%%%%%%%%%%%%%%%%%%%%%%%%%%%%%%%%%%%%%%%%%%%%%%%%%%%%%%%%%%%%%%%%%%%%

\renewcommand\theequation{A\arabic{equation}}

%\iffalse

\section{Similarity between partial transpose and exchange of vortices in chiral superconductors}
\label{app:partial transpose and exchange}

In this appendix, we briefly review the definition of partial transpose for fermions and show that the partial transpose is identical to the braiding operator of vortices in a chiral $p_x+ip_y$ superconductor. Let ${\cal H}$ be a fermionic Fock space  generated by $N$ local fermionic modes $f_j$, $j=1,\cdots,N$. 
The Hilbert space is spanned by $\ket{n_1,n_2,\cdots,n_N}$ which is a string of occupation numbers $n_j=0,1$. The Majorana (real) fermion operators are defined by
\begin{align}
c_{2j-1} \defineas f^{\dag}_j+f_j, \quad 
c_{2j} \defineas i(f_j-f_j^{\dag}), \quad 
j=1, \dots, N. 
\label{eq:real_fermion}
\end{align}
These operators satisfy the Clifford algebra $\{c_j, c_k \} = 2 \delta_{jk}$. Any operator $X$ acting on $\Hi$ can be expressed in terms of a polynomial of $c_j$'s,
\begin{align}
X = \sum_{k=1}^{2N} \sum_{p_1<p_2 \cdots <p_k} X_{p_1 \cdots p_k} c_{p_1} \cdots c_{p_k}, 
\label{eq:op_expand}
\end{align}
where $X_{p_1 \dots p_k}$ are complex coefficients which are fully antisymmetric under odd permutations of $\{1, \dots, k\}$. A density matrix is a Hermitian operator and commutes with the total fermion-number parity operator, $[\rho,(-1)^F]=0$ where $F=\sum_j f_j^\dag f_j$. The latter constraint implies that $\rho$ only contains monomials with even number of Majorana operators, i.e., $k$ is always even.

To define the bipartite entanglement, we consider decomposing the Hilbert space into $\Hi_1\otimes \Hi_2$. A generic density matrix on $\Hi_1\otimes \Hi_2$ can be expanded in the Majorana operators as
\begin{align} \label{eq:density_bipartite}
\rho = \sum_{k_1,k_2}^{k_1+k_2 = {\rm even}} \rho_{p_1 \cdots p_{k_1}, q_1 \cdots q_{k_2}} a_{p_1} \cdots a_{p_{k_1}} b_{q_1} \cdots b_{q_{k_2}},
\end{align}
where
$\{ a_j \}$ and $\{ b_j \}$ are Majorana operators acting on $\mathcal{H}_1$
and $\mathcal{H}_2$, respectively.
Our definition of the partial transpose for fermions is given by~\cite{Shap_pTR,Shiozaki_antiunitary}
\begin{align}
 \rT \defineas \sum_{k_1,k_2}^{k_1+k_2 = {\rm even}} \rho_{p_1 \cdots p_{k_1}, q_1 \cdots q_{k_2}} i^{k_1} a_{p_1} \cdots a_{p_{k_1}} b_{q_1} \cdots b_{q_{k_2}},
\label{eq:fermion_pt}
\end{align}
and similarly for $ \rho^{T_B}$. 

Let us now consider a simple example of $N=2$ complex fermions which share a Majorana dimer. 
 This state is described by the following density matrix
\begin{align}
\rho=\frac{1}{2}(1+i c_2 c_3),
\end{align}
where $c_i$'s are defined in (\ref{eq:real_fermion}). This operator is simply a projector into the subspace where $ic_2 c_3=1$.
Using the definition (\ref{eq:fermion_pt}), the partial transpose of the above density matrix is given by
\begin{align}\label{eq:pT_Maj_dimer}
\rho^{T_A}=\frac{1}{2}(1- c_2 c_3).
\end{align}

We now compare the partial transpose with the exchange  statistics of vortices.
To this end, we consider vortices in a spinless chiral superconductor. It is well-known that there exists a single Majorana bound state attached to each $\pi$-flux vortex~\cite{Volovik1999,Kopnin1991,Read2000}. The Majorana bound states can be described by operators $\gamma_i$ that mutually anti-commute and square to $+1$, i.e., they obey a Clifford algebra analogous to the ordinary Majorana operators defined above. 
Furthermore, two vortices can be combined, and the two localized Majorana bound states form a single complex fermion state which can be occupied or un-occupied. Hence, two vortices give a degeneracy of $2$. Similarly, we will get a $2^{N}$ ground state degeneracy for a collection of $2N$ vortices~\cite{Moore1991,Nayak1996}. It is important to remember that each vortex $\gamma_i$ is accompanied by a $\pi$-flux which can be represented by a branch cut. The branch cut is there to take into account the fact that a fermion picks up a $-1$ phase factor upon traversing around the vortex.

Let us now find the braiding operator of two vortices following the Refs.~\cite{Ivanov2001,Bernevig2013}. 
Exchanging two vortices $\gamma_i$ and $\gamma_j$, denoted by $T_{ij}$, is an adiabatic process and we are only interested in the unitary operator of the outcome. As a result, we get 
\begin{align}
\gamma_i\to \gamma_j,\qquad
\gamma_j\to -\gamma_i, \qquad
\gamma_k \to \gamma_k, 
\label{eq:majexchange}
\end{align} 
where $k\neq i, j$. 
  One can construct a representation of this exchange process on the Hilbert space by finding $\tau(T_{ij})$ such that $\tau(T_{ij})\gamma_a\tau^{-1}(T_{ij})=T_{ij}(\gamma_a).$ Such a representation is given by
\begin{align}
\tau(T_{ij})=\exp\left(\frac{\pi}{4}\gamma_{j}\gamma_{i}\right)=\frac{1}{\sqrt{2}}\left(1-\gamma_{i}\gamma_{j}\right).
\end{align}
It is easy to check that $\tau \gamma_i\tau^{-1}=\gamma_j$, $\tau \gamma_j\tau^{-1}=-\gamma_i$, and $\tau \gamma_k\tau^{-1}=\gamma_k$ for $k\neq i,j$.
Notice the similarity between the exchange operator above and the partial transpose in Eq.~(\ref{eq:pT_Maj_dimer}). This further supports the idea that partial transpose can be implemented by the exchange operator (or half braid) in anyonic systems.

%\fi

%%%%%%%%%%%%%%%%%%%%%%%%%%%%%%%%%%%%%%%%%%%%%%%
%%%%%%%%%%%%%%%%%%%%%%%%%%%%%%%%%%%%%%%%%%%%%%%

\section{Review of anyon diagrams}
\label{app:diagrams}

\renewcommand\theequation{B\arabic{equation}}

In this appendix, we discuss some basics of the modular tensor category which were used in the main text. Consult Refs.~\cite{Bonderson_thesis2012,BondersonShtengel2008,KnappBonderson2017} for a survey on this topic.
An anyon model $\mathcal{C}$ consists of a set of anyons which are labeled as $\{ a, b, c, \cdots \}$ and obey a commutative associative fusion algebra:
\begin{align}
    a\otimes b = \bigoplus_c N_{ab}^c c,
\end{align}
where $N_{ab}^c=N_{ba}^c$ are non-negative integers which give the number of ways anyons $a$ and $b$ can fuse to anyon $c$. An anyon $a$ is non-Abelian if $\sum_c N_{ab}^c>1$ for some $b$, and Abelian otherwise.

Physics of anyons as point-like excitations in 2D topologically ordered phases imposes certain constraints on the fusion algebra. There must exist a unique {vacuum} anyon $I$ such that $N_{aI}^c=\delta_{ac}$, and each anyon $a$ must have a unique conjugate charge or anti-particle $\bar{a}$ such that $N_{ab}^I=\delta_{b\bar{a}}$. The fusion coefficients also satisfy the following relation
\begin{align}\label{eq:Nabc}
    d_a d_b = \sum_c N_{ab}^c d_c,
\end{align}
where $d_a$, the quantum dimension of $a$, is the largest eigenvalue of the fusion matrix $N_a$ (when $N_{ab}^c$ is viewed as matrix elements $[N_a]_{bc}$). $d_a>1$ implies non-Abelian anyons, while $d_a=1$ implies Abelian anyons.

The fusion rules provide a direct
way to define the Hilbert space of anyons and states therein. In this paper, we use the diagrammatic formalism to denote states and operators. 
The building blocks of the anyonic Hilbert space is the space $V_c^{ab}$ of two anyons $a$ and $b$ with definite total charge $c$, which is spanned by the ket states
\begin{align}
    \ket{a,b;c,\mu}=\left(\frac{d_c}{d_a d_b}\right)^{1/4}\
    \begin{pspicture}[shift=-0.39](0,0)(1.2,1)
        \scriptsize
        \psline[ArrowInside=->](0.5,0.5)(0,1)\rput(0.1,0.7){$a$}
        \psline[ArrowInside=->](0.5,0.5)(1,1)\rput(0.95,0.7){$b$}
        \psline[ArrowInside=->](0.5,0)(0.5,0.5)\rput(0.65,0.2){$c$}
        \rput(0.65,0.45){$\mu$}
    \end{pspicture},
\end{align}
where $\mu=1,\dots,N_{ab}^c$. The dual space $V_{ab}^c$ is spanned by the bra states
\begin{align}
    \bra{a,b;c,\mu}=\left(\frac{d_c}{d_a d_b}\right)^{1/4}
    \begin{pspicture}[shift=-0.38](0,0)(1.3,1)
        \scriptsize
        \psline[ArrowInside=->](0,0)(0.5,0.5)\rput(0.1,0.32){$a$}
        \psline[ArrowInside=->](1,0)(0.5,0.5)\rput(0.95,0.32){$b$}
        \psline[ArrowInside=->](0.5,0.5)(0.5,1)\rput(0.65,0.8){$c$}
        \rput(0.65,0.55){$\mu$}
    \end{pspicture}.
\end{align}

Inner products can be evaluated by stacking the diagrams. For instance, the orthonormality condition
\begin{align}
    \langle a',b';c',\mu'|a,b;c,\mu\rangle=\delta_{aa'}\delta_{bb'}\delta_{cc'}\delta_{\mu\mu'} \openone_{c}
\end{align}
can be expressed as
\begin{align}
\label{eq:orthogonality}
    \left(\frac{d_c^2}{d_a d_b d_{a'} d_{b'}}\right)^{1/4}
    \begin{pspicture}[shift=-0.9](-0.2,0)(1.2,2)
        \scriptsize
        \psline[ArrowInside=->](0.5,0.5)(0,1)\rput(0.1,0.75){$a$}
        \psline[ArrowInside=->](0.5,0.5)(1,1)\rput(0.95,0.75){$b$}
        \psline[ArrowInside=->](0.5,0)(0.5,0.5)\rput(0.65,0.2){$c$}
        \rput(0.65,0.45){$\mu$}
        \rput(0,1){
        \psline[ArrowInside=->](0,0)(0.5,0.5)\rput(0.1,0.25){$a'$}
        \psline[ArrowInside=->](1,0)(0.5,0.5)\rput(1,0.25){$b'$}
        \psline[ArrowInside=->](0.5,0.5)(0.5,1)\rput(0.7,0.8){$c'$}
        \rput(0.7,0.55){$\mu'$}}
    \end{pspicture}
    =
    \delta_{a,a'}\delta_{b,b'}\delta_{c,c'}\delta_{\mu,\mu'}
    \begin{pspicture}[shift=-0.9](-0.2,0)(0.4,2)
        \scriptsize
        \psline[ArrowInside=->](0,0)(0,2)\rput(0.2,1){$c$}
    \end{pspicture}.
\end{align}
Note that in the diagramatic notation, $\delta_{cc'}$ is to enforce the conservation of anyonic charge. More complicated diagrams can be evaluated similarly.
In particular, we can choose $c=I$ 
and use the above relation to determine the quantum dimension $d_a$. Diagrammatically, this corresponds to
\begin{align}
d_a=
    \begin{pspicture}[shift=-0.35](-0.08,0.25)(1.55,1.25)
    \scriptsize
        \psarc[linecolor=black,arrows=<-,
    arrowinset=0.15] (0.8,0.7){0.5}{165}{363}
        \psarc[linecolor=black,border=0pt]
(0.8,0.7){0.5}{0}{170}
        \rput[bl]{0}(-0.03,0.55){$a$}
    \end{pspicture},
\end{align}
where we use the fact that $d_I=1$. 
We also note that $d_a=d_{\bar{a}}$.
In general, a diagram with open anyon worldlines at the top and bottom represents a state in a Hilbert space that depends on the number and types of open anyon worldlines. A diagram without open worldlines represents an amplitude or a complex number.

The space $V_{a b}^{a' b'}$ of operators acting on anyons $a$ and $b$ can be constructed as $V^{a'b'}_{ab}=\bigoplus_c V^c_{ab}\otimes V_c^{ab}$,
which is spanned by
\begin{align}
    \ket{a',b',\alpha';c}&\bra{a,b,\alpha;c}
    = \left( \frac{d_c^2}{d_a d_b d_{a'} d_{b'}} \right)^{1/4}    \begin{pspicture}[shift=-0.6](-0.2,0)(1.2,1.5)
        \scriptsize
        \psline[ArrowInside=->](0,0)(0.5,0.5)\rput(0,0.25){$a$}
        \psline[ArrowInside=->](1,0)(0.5,0.5)\rput(1,0.25){$b$}
        \rput(0.7,0.5){$\alpha$}
        \psline[ArrowInside=->](0.5,0.5)(0.5,1)\rput(0.65,0.75){$c$}
        \rput(0.7,1){$\alpha'$}
        \psline[ArrowInside=->](0.5,1)(0,1.5)\rput(0,1.25){$a'$}
        \psline[ArrowInside=->](0.5,1)(1,1.5)\rput(1,1.25){$b'$}
    \end{pspicture}.
\end{align}
For example, the identity operator for a pair of anyons $a$ and $b$ is
\begin{align}
    \openone_{ab}=\sum_{c,\mu} \ket{a,b;c,\mu}\bra{a,b;c,\mu},
\end{align}
or, diagramatically,
\begin{align}
\label{eq:identity-op}
    \begin{pspicture}[shift=-0.6](-0.2,0)(1.4,1.5)
        \scriptsize
        \psline[ArrowInside=->](0,0)(0,1.5)\rput(0.2,0.75){$a$}
        \psline[ArrowInside=->](1,0)(1,1.5)\rput(1.2,0.75){$b$}
    \end{pspicture}
    =\sum_{c,\mu} \sqrt{\frac{d_c}{d_a d_b}}
    \begin{pspicture}[shift=-0.6](-0.2,0)(1.2,1.5)
        \scriptsize
        \psline[ArrowInside=->](0,0)(0.5,0.5)\rput(0,0.25){$a$}
        \psline[ArrowInside=->](1,0)(0.5,0.5)\rput(1,0.25){$b$}
        \rput(0.7,0.5){$\mu$}
        \psline[ArrowInside=->](0.5,0.5)(0.5,1)\rput(0.65,0.75){$c$}
        \rput(0.7,1){$\mu$}
        \psline[ArrowInside=->](0.5,1)(0,1.5)\rput(0,1.25){$a$}
        \psline[ArrowInside=->](0.5,1)(1,1.5)\rput(1,1.25){$b$}
    \end{pspicture}.
\end{align}

Larger Hilbert spaces are constructed by a fusion tree. For instance, the space of three anyons $a$, $b$, and $c$ with definite total charge $d$, denoted as $V^{abc}_d$, can be constructed by a sum of tensor products
$V^{abc}_d\cong\bigoplus_e V^{ab}_e\otimes V^{ec}_d$,
which is spanned by
\begin{align}
    \ket{a,b;e,\mu}\ket{e,c;d,\nu}=\left(\frac{d_d}{d_a d_b d_c}\right)^{1/4}\
    \begin{pspicture}[shift=-0.7](0,0)(2.1,1.5)
        \scriptsize
        \psline[ArrowInside=->](0.5,1)(0,1.5)\rput(0.1,1.2){$a$}
        \psline[ArrowInside=->](0.5,1)(1,1.5)\rput(0.9,1.2){$b$}
        \psline[ArrowInside=->](1,0.5)(2,1.5)\rput(1.65,0.9){$c$}
        \psline[ArrowInside=->](1,0)(1,0.5)\rput(1.15,0.25){$d$}
        \psline[ArrowInside=->](1,0.5)(0.5,1)\rput(0.6,0.7){$e$}
        \rput(0.38,0.93){$\mu$}
        \rput(1.15,0.5){$\nu$}
    \end{pspicture},
\end{align}
where $\mu=1,\dots,N_{ab}^e$, $\nu=1,\dots,N_{ec}^d$, upon introducing the intermediate anyon $e$. The space $V^{abc}_d$ can also be constructed by another decomposition,
$V^{abc}_d\cong\bigoplus_e V^{bc}_e\otimes V^{ae}_d$.
The two constructions are isomorphic, and their basis vectors are related by an $F$-move:
\begin{align}
    \begin{pspicture}[shift=-0.6](0,0)(2,1.5)
        \scriptsize
        \psline[ArrowInside=->](1,0.5)(2,1.5)\rput(1.65,0.9){$c$}
        \psline[ArrowInside=->](0.5,1)(0,1.5)\rput(0.1,1.2){$a$}
        \psline[ArrowInside=->](0.5,1)(1,1.5)\rput(0.9,1.2){$b$}
        \psline[ArrowInside=->](1,0)(1,0.5)\rput(1.15,0.25){$d$}
        \psline[ArrowInside=->](1,0.5)(0.5,1)\rput(0.6,0.7){$e$}
        \rput(0.35,0.93){$\mu$}
        \rput(1.15,0.45){$\nu$}
    \end{pspicture}
    \defineas \sum_{f,\alpha,\beta} \left[F^{abc}_d\right]_{(e,\mu,\nu)(f,\alpha,\beta)}
    \begin{pspicture}[shift=-0.6](0.2,0)(2,1.5)
        \scriptsize
        \psline[ArrowInside=->](1.5,1)(1,1.5)\rput(1.1,1.2){$b$}
        \psline[ArrowInside=->](1.5,1)(2,1.5)\rput(1.9,1.2){$c$}
        \psline[ArrowInside=->](1,0.5)(0,1.5)\rput(0.35,0.9){$a$}
        \psline[ArrowInside=->](1,0)(1,0.5)\rput(1.15,0.25){$d$}
        \psline[ArrowInside=->](1,0.5)(1.5,1)\rput(1.4,0.7){$f$}
        \rput(1.65,0.93){$\alpha$}
        \rput(1.15,0.45){$\beta$}
    \end{pspicture},
\end{align}
where the $F$-symbols $F^{abc}_d$ are unitary matrices which must satisfy a consistency condition, the so-called Pentagon equations.

In general, the space $V^{a_1\dots a_n}_c$ of anyons $a_1$, \dots, $a_n$ with definite combined charge $c$ can be constructed as
\begin{align}
    V^{a_1\dots a_n}_c\cong\bigoplus_{\vec{b}} V^{a_1 a_2}_{b_2} \otimes V^{b_2 a_3}_{b_3}\otimes \cdots \otimes V^{b_{n-1} a_n}_c,
\end{align}
which is spanned by
\begin{align}
    \ket{\vec{a},\vec{b},\vec{\alpha};c}&=\ket{a_1,a_2;b_2,\alpha_2}\cdots\ket{b_{{n-1}},a_n;c,\alpha_n} \notag \\
    &=\left(\frac{d_{c}}{d_{\vec{a}}}\right)^{1/4}
    \begin{pspicture}[shift=-0.7](-0.2,0)(2.2,1.8)
        \scriptsize
        \psline[ArrowInside=->](1,0.5)(2,1.5)\rput(1.8,1){$a_n$}
        \psline[ArrowInside=->](0.5,1)(1,1.5)\rput(1,1.25){$a_2$}
        \psline[ArrowInside=->](0.5,1)(0,1.5)\rput(0.05,1.25){$a_1$}
        \psline[ArrowInside=->](1,0)(1,0.5)\rput(1.2,0.2){$c$}
        \psline[ArrowInside=->](0.7,0.8)(0.5,1)\rput(0.5,0.75){$b_2$}
        \psline[ArrowInside=->](1,0.5)(0.8,0.7)\rput(0.6,0.5){$b_{n-1}$}
        \rput(0.25,1){$\alpha_2$}
        \rput(1.3,0.5){$\alpha_n$}
        \rput(0.75,0.75){.}
        \rput(1.5,1.5){\dots}
    \end{pspicture},
\end{align}
where $\vec{b}$ and $\vec{\alpha}$ take values that are allowed by fusion and we also define
\begin{align}
    d_{\vec{a}} \defineas d_{a_1}\cdots d_{a_n}=\sum_c N_{a_1 \dots a_n}^c d_c.
\end{align}

Similarly, the space $V_{a_1\dots a_n}^{a_1'\dots a_n'}$ of operators acting on anyons $a_1$, \dots, $a_n$ can be constructed as
\begin{align}
    V_{a_1\dots a_n}^{a_1'\dots a_n'}=\bigoplus_c V^c_{a_1\dots a_n}\otimes V^{a_1'\dots a_n'}_c,
\end{align}
which is spanned by
\begin{align}
    \ket{\vec{a}',\vec{b}',\vec{\alpha}';c}&\bra{\vec{a},\vec{b},\vec{\alpha};c} =\left(\frac{d_c^2}{d_{\vec{a}}d_{\vec{a}'}}\right)^{1/4}
    \begin{pspicture}[shift=-1.5](-0.2,-1.5)(2.2,1.5)
        \scriptsize
        \psline[ArrowInside=->](0.5,1)(0,1.5)\rput(0,1.25){$a'_1$}
        \psline[ArrowInside=->](0.5,1)(1,1.5)\rput(1,1.25){$a'_2$}
        \psline[ArrowInside=->](1,0.5)(2,1.5)\rput(1.8,1){$a'_n$}
        \psline[ArrowInside=->](1,0)(1,0.5)\rput(1.15,0.25){$c$}
        \psline[ArrowInside=->](0.7,0.8)(0.5,1)\rput(0.5,0.75){$b'_2$}
        \psline[ArrowInside=->](1,0.5)(0.8,0.7)\rput(0.6,0.5){$b'_{n-1}$}
        \rput(0.2,1){$\alpha'_2$}
        \rput(1.3,0.5){$\alpha'_n$}
        \rput(0.75,0.75){.}
        \rput(1.5,1.5){\dots}
        \rput(0,0.5){
        \psline[ArrowInside=->](0,-1.5)(0.5,-1)\rput(0.05,-1.25){$a_1$}
        \psline[ArrowInside=->](1,-1.5)(0.5,-1)\rput(1,-1.25){$a_2$}
        \psline[ArrowInside=->](2,-1.5)(1,-0.5)\rput(1.8,-1){$a_n$}
        \psline[ArrowInside=->](0.5,-1)(0.7,-0.8)\rput(0.5,-0.75){$b_2$}
        \psline[ArrowInside=->](0.8,-0.7)(1,-0.5)\rput(0.6,-0.45){$b_{n-1}$}
        \rput(0.25,-1){$\alpha_2$}
        \rput(1.3,-0.5){$\alpha_n$}
        \rput(0.75,-0.75){.}
        \rput(1.5,-1.5){\dots}}
    \end{pspicture},
\end{align}
where $\vec{b}$, $\vec{\alpha}$, $\vec{b}'$, and $\vec{\alpha}'$ take values that are allowed by fusion. 
The above definition of operators which act on the Hilbert space of multiple anyons is the basis of the next part, where we discuss a model Hamiltonian in terms of a fusion tree.

We also use the $A$-moves which are associated with bending the wordlines as follows
\begin{align}
    \label{eq:Amove}
    \begin{pspicture}[shift=-0.39](0,0)(1.2,1)
        \scriptsize
        \psline[ArrowInside=->](0.5,0.5)(0,1)\rput(0.1,0.7){$a$}
        \psline[ArrowInside=-<](0,1)(-0.5,0.5) \rput(-0.55,0.7){$\bar a$}
        \psline[ArrowInside=->](0.5,0.5)(1,1)\rput(0.95,0.7){$b$}
        \psline[ArrowInside=->](0.5,0)(0.5,0.5)\rput(0.65,0.2){$c$}
        \rput(0.65,0.45){$\mu$}
    \end{pspicture}
    \defineas
    \sum_{\nu} [A^{ab}_c]_{\mu\nu}
    \begin{pspicture}[shift=-0.38](-0.1,0)(1.3,1)
        \scriptsize
        \psline[ArrowInside=->](0,0)(0.5,0.5)\rput(0.1,0.32){$\bar a$}
        \psline[ArrowInside=->](1,0)(0.5,0.5)\rput(0.95,0.32){$c$}
        \psline[ArrowInside=->](0.5,0.5)(0.5,1)\rput(0.65,0.8){$b$}
        \rput(0.65,0.55){$\nu$}
    \end{pspicture},
\end{align}
where the $A$-symbols are given in terms of the $F$-symbols
\begin{align}
    \label{eq:Asymbol}
    [A^{ab}_c]_{\mu\nu}= \sqrt{\frac{d_a d_b}{d_c}} \varkappa_a^\ast [F^{\bar a a b}_b]^\ast_{1,(c,\mu,\nu)}.
\end{align}
One can similarly define the $B$-moves by bending the anyon $b$'s worldline.

Another useful operation that we utilize in this paper, is the braiding exchange operator for a pair of anyons,
\begin{align}
R^{ab}=\sum_{c,\mu} [R^{ab}_c]_{\mu\nu} \ket{a,b;c,\mu}\bra{b,a;c,\nu},
\end{align}
or, diagramatically,
\begin{align}
    R^{ab}=\begin{pspicture}[shift=-0.6](-0.2,0)(1.2,1.5)
        \scriptsize
        \psline(1,0)(0,1.5)
        \psline[border=2pt](0,0)(1,1.5)
        \psline[ArrowInside=->](1,0)(0.66,0.5)\rput(1,0.3){$a$}
        \psline[ArrowInside=->](0,0)(0.33,0.5)\rput(0,0.3){$b$}
    \end{pspicture}
    \defineas \sum_{c,\mu,\nu} \sqrt{\frac{d_c}{d_a d_b}} [R^{ab}_c]_{\mu\nu}
    \begin{pspicture}[shift=-0.6](-0.2,0)(1.2,1.5)
        \scriptsize
        \psline[ArrowInside=->](0,0)(0.5,0.5)\rput(0,0.25){$b$}
        \psline[ArrowInside=->](1,0)(0.5,0.5)\rput(1,0.25){$a$}
        \rput(0.7,0.5){$\nu$}
        \psline[ArrowInside=->](0.5,0.5)(0.5,1)\rput(0.65,0.75){$c$}
        \rput(0.7,1){$\mu$}
        \psline[ArrowInside=->](0.5,1)(0,1.5)\rput(0,1.25){$a$}
        \psline[ArrowInside=->](0.5,1)(1,1.5)\rput(1,1.25){$b$}
    \end{pspicture},
\end{align}
where the $R$-symbols $R^{ab}_c$ are unitary matrices that must satisfy the Hexagon consistency equations. The above operation is usually called a counterclockwise braiding exchange. We may also have a clockwise braiding exchange as,
\begin{align}
    (R^{ab})^{-1}=\begin{pspicture}[shift=-0.6](-0.2,0)(1.2,1.5)
        \scriptsize
        \psline(0,0)(1,1.5)
        \psline[border=2pt](1,0)(0,1.5)
        \psline[ArrowInside=->](0,0)(0.33,0.5)\rput(0,0.3){$a$}
        \psline[ArrowInside=->](1,0)(0.66,0.5)\rput(1,0.3){$b$}
    \end{pspicture}.
\end{align}
An important quantity derived from braiding exchange is the topological twist (or topological spin) of charge $a$
\begin{align}
\theta_a=\theta _{\bar{a}}
=\sum_{c,\mu } \frac{d_c}{d_a}\left[ R_c^{aa}\right] _{\mu \mu }
= \frac{1}{d_a}
\begin{pspicture}[shift=-0.5](-1.3,-0.6)(1.3,0.6)
\scriptsize
   %%% Arcs
  \psarc(0.7071,0.0){0.5}{-135}{135}
  \psarc(-0.7071,0.0){0.5}{45}{315}
  %%% Straight segments
  \psline(-0.3536,0.3536)(0.3536,-0.3536)
  \psline[border=2.3pt](-0.3536,-0.3536)(0.3536,0.3536)
  \psline[border=2.3pt]{->}(-0.3536,-0.3536)(0.0,0.0)
  \rput[bl]{0}(-0.2,-0.5){$a$}
 \end{pspicture},
\end{align}
which is a root of unity. 
 
A natural way to define anyonic density matrices is by using partial tracing which results in a reduced density matrix of a subsystem.
We require the ordinary trace of an operator to be the sum of its diagonal elements, e.g.,
\begin{align}\label{eq:egtrace}
    \Tr(\ket{a',b';c,\mu'}\bra{a,b;c,\mu}) \defineas \delta_{aa'}\delta_{bb'}\delta_{\mu\mu'} .
\end{align}
Similarly, the anyonic trace in the diagrammatic representation denoted by the \emph{quantum trace} $\aTr$, (also called the \emph{anyonic trace})
is realized by connecting the outgoing and incoming anyon lines.
Diagrams which contain loops that connect two lines labeled by different topological charges are identically zero.
The partial anyonic trace is also obtained by connecting only the outgoing and incoming lines of the anyons which we want to trace over.
The anyonic trace in the diagrammatic representation is denoted by the \emph{quantum trace} $\aTr$, (also called the \emph{anyonic trace}),
\begin{align}
&\aTr\Bigg(\left(\frac{d_c^2}{d_a d_b d_{a'} d_{b'}}\right)^{1/4}
    \begin{pspicture}[shift=-1.2](-0.2,-0.5)(1.2,1.5)
        \scriptsize
        \psline[ArrowInside=->](0,0)(0.5,0.5)\rput(0,0.25){$a$}
        \psline[ArrowInside=->](1,0)(0.5,0.5)\rput(1,0.25){$b$}
        \rput(0.7,0.5){$\mu$}
        \psline[ArrowInside=->](0.5,0.5)(0.5,1)\rput(0.65,0.75){$c$}
        \rput(0.7,1){$\mu'$}
        \psline[ArrowInside=->](0.5,1)(0,1.5)\rput(0,1.25){$a'$}
        \psline[ArrowInside=->](0.5,1)(1,1.5)\rput(1,1.25){$b'$}
    \end{pspicture}
    \Bigg) \nonumber \\
   & \defineas \left(\frac{d_c^2}{d_a d_b d_{a'} d_{b'}}\right)^{1/4}
    \begin{pspicture}[shift=-1.2](-0.2,-0.6)(1.8,1.7)
        \scriptsize
        \psline[ArrowInside=->](0,0)(0.5,0.5)\rput(0,0.25){$a$}
        \psline[ArrowInside=->](1,0)(0.5,0.5)\rput(1,0.25){$b$}
        \rput(0.7,0.5){$\mu$}
        \psline[ArrowInside=->](0.5,0.5)(0.5,1)\rput(0.65,0.75){$c$}
        \rput(0.7,1){$\mu'$}
        \psline[ArrowInside=->](0.5,1)(0,1.5)(0.75,2.25)(1.75,1.25)(1.75,0.25)(0.75,-0.75)(0,0)\rput(0,1.25){$a'$}
        \psline[ArrowInside=->](0.5,1)(1,1.5)(1.5,1)(1.5,0.5)(1,0)\rput(1,1.25){$b'$}
    \end{pspicture}
 \\
&= d_c \delta_{a,a'}\delta_{b,b'}\delta_{\mu,\mu'},
\end{align}
We should note that for the Abelian fusion channels $d_c=1$ and the ordinary trace and anyonic trace are identical.
%diagrams which contain loops that connect two lines labeled by different topological charge values are identically zero, due to the orthogonality condition.

The partial  anyonic trace 
is also obtained by connecting only the outgoing and incoming lines of the anyons which we want to trace over. For instance,
\begin{align}
    &\aTr_b\Bigg(\left(\frac{d_c^2}{d_a d_b d_{a'} d_{b'}}\right)^{1/4}
    \begin{pspicture}[shift=-0.6](-0.2,0)(1.2,1.5)
        \scriptsize
        \psline[ArrowInside=->](0,0)(0.5,0.5)\rput(0,0.25){$a$}
        \psline[ArrowInside=->](1,0)(0.5,0.5)\rput(1,0.25){$b$}
        \rput(0.7,0.5){$\mu$}
        \psline[ArrowInside=->](0.5,0.5)(0.5,1)\rput(0.65,0.75){$c$}
        \rput(0.7,1){$\mu'$}
        \psline[ArrowInside=->](0.5,1)(0,1.5)\rput(0,1.25){$a'$}
        \psline[ArrowInside=->](0.5,1)(1,1.5)\rput(1,1.25){$b'$}
    \end{pspicture}
    \Bigg) \nonumber \\
    & \defineas \left(\frac{d_c^2}{d_a d_b d_{a'} d_{b'}}\right)^{1/4}
    \begin{pspicture}[shift=-0.6](-0.2,0)(1.7,1.5)
        \scriptsize
        \psline[ArrowInside=->](0,0)(0.5,0.5)\rput(0,0.25){$a$}
        \psline[ArrowInside=->](1,0)(0.5,0.5)\rput(1,0.25){$b$}
        \rput(0.7,0.5){$\mu$}
        \psline[ArrowInside=->](0.5,0.5)(0.5,1)\rput(0.65,0.75){$c$}
        \rput(0.7,1){$\mu'$}
        \psline[ArrowInside=->](0.5,1)(0,1.5)\rput(0,1.25){$a'$}
        \psline[ArrowInside=->](0.5,1)(1,1.5)(1.5,1)(1.5,0.5)(1,0)\rput(1,1.25){$b'$}
    \end{pspicture}  \\
    &=\frac{d_c}{d_a} \delta_{aa'}\delta_{bb'}\delta_{\mu\mu'}
    \begin{pspicture}[shift=-0.6](-0.2,0)(0.4,1.5)
        \scriptsize
        \psline[ArrowInside=->](0,0)(0,1.5)\rput(0.2,0.75){$a$}
    \end{pspicture}.
\end{align}

In the diagrammatic representation, a generic operator $X\in V^{a_1 \dots a_n}_{a'_1 \dots a'_n}$ is shown as
\begin{align}
\label{eq:operator}
X \defineas
\psscalebox{.7}{
 \pspicture[shift=-0.8](-1,-0.9)(1,0.8)
  \small
%%%%% Box:
  \psframe[linewidth=0.9pt,linecolor=black,border=0](-0.8,-0.5)(0.8,0.5)
  \rput[bl]{0}(-0.15,-0.1){$X$}
  \rput[bl]{0}(-0.22,0.7){$\mathbf{\ldots}$}
  \rput[bl]{0}(-0.22,-0.75){$\mathbf{\ldots}$}
%%%%% Line connections:
  \psset{linewidth=0.9pt,linecolor=black,arrowscale=1.5,arrowinset=0.15}
  \psline(0.6,0.5)(0.6,1)   \rput(0.6,1.2){$a_n'$}
  \psline(-0.6,0.5)(-0.6,1) \rput(-0.6,1.2){$a_1'$}
  \psline(0.6,-0.5)(0.6,-1) \rput(0.6,-1.2){$a_n$}
  \psline(-0.6,-0.5)(-0.6,-1) \rput(-0.6,-1.2){$a_1$}
%%%%% Arrows:
  \psline{->}(0.6,0.5)(0.6,0.9)
  \psline{->}(-0.6,0.5)(-0.6,0.9)
  \psline{-<}(0.6,-0.5)(0.6,-0.9)
  \psline{-<}(-0.6,-0.5)(-0.6,-0.9)
\endpspicture
},
\end{align}
and the corresponding anyonic trace is given by
\begin{align}
\label{eq:trace_illustrate}
\aTr X =
\aTr
\left[
\psscalebox{.7}{
 \pspicture[shift=-1.0](-1,-1.1)(1,1.1)
  \small
%%%%% Box:
  \psframe[linewidth=0.9pt,linecolor=black,border=0](-0.8,-0.5)(0.8,0.5)
  \rput[bl]{0}(-0.15,-0.1){$X$}
  \rput[bl]{0}(-0.22,0.7){$\mathbf{\ldots}$}
  \rput[bl]{0}(-0.22,-0.75){$\mathbf{\ldots}$}
%%%%% Line connections:
  \psset{linewidth=0.9pt,linecolor=black,arrowscale=1.5,arrowinset=0.15}
  \psline(0.6,0.5)(0.6,1)
  \psline(-0.6,0.5)(-0.6,1)
  \psline(0.6,-0.5)(0.6,-1)
  \psline(-0.6,-0.5)(-0.6,-1)
%%%%% Arrows:
  \psline{->}(0.6,0.5)(0.6,0.9)
  \psline{->}(-0.6,0.5)(-0.6,0.9)
  \psline{-<}(0.6,-0.5)(0.6,-0.9)
  \psline{-<}(-0.6,-0.5)(-0.6,-0.9)
\endpspicture
}
\right]
= \sum_{a_1 , \ldots , a_n}
\psscalebox{.7}{
\pspicture[shift=-1.1](-1.0,-1.2)(2.3,1.1)
  \small
%%%%% Labels:
  \psframe[linewidth=0.9pt,linecolor=black,border=0](-0.8,-0.5)(0.8,0.5)
  \rput[bl]{0}(-0.15,-0.1){$X$}
  \rput[bl]{0}(-0.4,0.7){$\mathbf{\ldots}$}
  \rput[bl]{0}(-0.22,-0.75){$\mathbf{\ldots}$}
  \rput[bl]{0}(1.52,0){$\mathbf{\ldots}$}
%%%%% Line connections:
  \psset{linewidth=0.9pt,linecolor=black,arrowscale=1.5,arrowinset=0.15}
%%%%% Arcs
  \psarc(1.0,0.5){0.4}{0}{180}
  \psarc(1.0,-0.5){0.4}{180}{360}
  \psarc(0,0.5){0.6}{90}{180}
  \psarc(0,-0.5){0.6}{180}{270}
  \psarc(1.5,0.5){0.6}{0}{90}
  \psarc(1.5,-0.5){0.6}{270}{360}
%%%%% Line connections:
  \psline(1.4,-0.5)(1.4,0.5)
  \psline(0,1.1)(1.5,1.1)
  \psline(0,-1.1)(1.5,-1.1)
  \psline(2.1,-0.5)(2.1,0.5)
%%%%% Arrows
  \psline{->}(1.4,0.2)(1.4,-0.1)
  \psline{->}(2.1,0.2)(2.1,-0.1)
%%%%% Leg Labels:
  \rput[bl](-1.07,0.6){$a_1$}
  \rput[bl](0.1,0.6){$a_n$}
 \endpspicture
}
.
\end{align}

Note that for a generic operator $X\in V^{a_1 \dots a_n}_{a'_1 \dots a'_n}$ the anyonic and ordinary traces
are related by
\begin{align}
\label{eq:an_trace}
    \aTr X &=\sum_c d_c\Tr [X]_c ,\\
    \Tr X &=\sum_c \frac{1}{d_c}\aTr [X]_c,
\end{align}
where $[X]_c = \Pi_c X \Pi_c\in V^{a_1 \dots a_n}_c\otimes V^c_{a'_1 \dots a'_n}$ is the projection of $X$ onto definite total fusion channel $c$, with $X=\sum_c [X]_c$.

%%%%%%%%%%%%%%%%%%%%%%%%%%%%%%%%%%%%%%%%%%%%%
%%%%%%%%%%%%%%%%%%%%%%%%%%%%%%%%%%%%%%%%%%%%%

\section{Lemmas for monotonicity}
\label{app:lemmas}

\renewcommand\theequation{C\arabic{equation}}

In this appendix, we prove two lemmas which are essential for proving the monotonicity under LOCC in Sec.\,\ref{sec:monotonicity}.

\vspace{1ex}

\noindent\textbf{1.~Appending ancilla:}
Appending an unentangled local ancilla $R$ must not change the entanglement measure.

This is modeled by the following process
\begin{align} 
\arho_{AB} \to ( \arho_{AB} \otimes \arho_R),
\end{align}
where we add an ancilla in an arbitrary mixed state, denoted by $\arho_R$, 
to our original system $\arho_{AB}$
and by local ancilla, we mean that the new global system $R \cup (A\cup B)$ is partitioned to  $A'=A\cup R$ and $B$. We need to show that
\begin{align} \label{eq:mono1_logN}
{\cal E}(\arho_{AB} \otimes \arho_R)= {\cal E}(\arho_{AB}).
\end{align}
A way to prove this starts by noticing that the partial transpose of $A'\cup B$ is now taken with respect to $A'$.
From the property (\ref{eq:ptrans_tensor}), we write
\begin{align}
    \norm{(\arho_{AB} \otimes \arho_R)^{T_{A'}}}_1 &= 
    \norm{\arho_{AB}^{T_{A'}} \otimes \arho_R^{T_{A'}}}_1 \nonumber \\
    &= \norm{\arho_{AB}^{T_{A}} \otimes \arho_R^{T}}_1 \nonumber \\
    &= \norm{\arho_{AB}^{T_{A}}}_1
    \label{eq:mono_ancilla}
\end{align}
where in the second line we simplify $T_{A'}$ into the partial transpose with respect to $A$ and full transpose for $A\cup B$ and $R$ subsystems, respectively, and in the third line we make use of the fact that the full transpose does not change the one norm as explained in Sec.\,\ref{sec:separability}.

\vspace{1ex}

\noindent\textbf{2.~Local projectors:}
 Application of local projective measurements does not increase the entanglement measure. A local projection operator which acts on the anyonic space $V^{a_1\cdots a_n}_{a_1 \cdots a_n}$ and projects into fusion  channel $c$ is given by
\begin{align}
    \Pi_{\vec{a}}^c
    =\left(\frac{d_c}{d_{\vec{a}}}\right)^{1/2} \sum_{\vec{e},\vec{\alpha}}
    \begin{pspicture}[shift=-1.5](-0.2,-1.5)(2.2,1.5)
        \scriptsize
        \psline[ArrowInside=->](0.5,1)(0,1.5)\rput(0,1.25){$a_1$}
        \psline[ArrowInside=->](0.5,1)(1,1.5)\rput(1,1.25){$a_2$}
        \psline[ArrowInside=->](1,0.5)(2,1.5)\rput(1.8,1){$a_n$}
        \psline[ArrowInside=->](1,0)(1,0.5)\rput(1.15,0.25){$c$}
        \psline[ArrowInside=->](0.7,0.8)(0.5,1)\rput(0.5,0.75){$e_2$}
        \psline[ArrowInside=->](1,0.5)(0.8,0.7)\rput(0.6,0.5){$e_{n-1}$}
        \rput(0.2,1){$\alpha_2$}
        \rput(1.3,0.5){$\alpha_n$}
        \rput(0.75,0.75){.}
        \rput(1.5,1.5){\dots}
        \rput(0,0.5){
        \psline[ArrowInside=->](0,-1.5)(0.5,-1)\rput(0.05,-1.25){$a_1$}
        \psline[ArrowInside=->](1,-1.5)(0.5,-1)\rput(1,-1.25){$a_2$}
        \psline[ArrowInside=->](2,-1.5)(1,-0.5)\rput(1.8,-1){$a_n$}
        \psline[ArrowInside=->](0.5,-1)(0.7,-0.8)\rput(0.5,-0.75){$e_2$}
        \psline[ArrowInside=->](0.8,-0.7)(1,-0.5)\rput(0.6,-0.45){$e_{n-1}$}
        \rput(0.25,-1){$\alpha_2$}
        \rput(1.3,-0.5){$\alpha_n$}
        \rput(0.75,-0.75){.}
        \rput(1.5,-1.5){\dots}}
    \end{pspicture}.
\end{align}
Note that the projection operators satisfy,
\begin{align}
    \label{eq:P_orthogonal}
    \Pi_{\vec{a}}^c \Pi_{\vec{a}}^{c'} &= \delta_{cc'} \Pi^c_{\vec{a}},\\
    \label{eq:P_complete}
    \sum_c \Pi^c_{\vec{a}} &= \openone_{\vec{a}}. 
\end{align}
The above identities are easy to show  diagrammatically.
 Now, consider two sets of orthogonal local projectors $\{\Pi^{(i)}_A\}$ and $\{\Pi^{(i)}_B\}$ on subsystems $A$ and $B$. Here, each $\{\Pi^{(i)}_A\}$ operator may include multiple local fusion channels. We consider an orthogonal measurement on the composite system using a product of local projection operators,
 \begin{align}
     \Pi^{(ij)}_{AB}= \Pi^{(i)}_A \otimes \Pi^{(j)}_B,
 \end{align}
which in turn form an orthocomplete set, i.e.,
 \begin{align}
     \sum_{i,j} \Pi^{(ij)}_{AB} &= \openone_{AB}, \\
     \Pi^{(ij)}_{AB} \Pi^{(i'j')}_{AB} &= \delta_{ii'} \delta_{jj'} \Pi^{(ij)}_{AB}.
 \end{align}
 We define locally projected density matrices by
\begin{align}
\arho_{AB}^{(ij)}=\frac{1}{r_{ij}} \Pi^{(ij)}_{AB} \arho_{AB} \Pi^{(ij)}_{AB} ,
\end{align}
where $r_{ij}=\aTr[ (\Pi^{(i)}_A \otimes \Pi^{(j)}_B) \arho_{AB} (\Pi^{(i)}_A\otimes \Pi^{(j)}_B)]$.
%The completeness of the set of projectors (\ref{eq:P_complete}) implies 
%\begin{align}
 %\Tr\left[ \sum_i  (\Pi^{e_i}_A \otimes \Pi^{f_i}_B) \arho_{AB} (\Pi^{e_i}_A\otimes \Pi^{f_i}_B) \right]=1,
%\end{align}
%which means $\sum_i r_i=1$.
 The monotonicity condition can then be written as
\begin{align} \label{eq:mono3_logN}
{\cal E}(\arho_{AB}) \geq \sum_{i,j}  r_{ij}  {\cal E}\left(\arho_{AB}^{(ij)}  \right).
\end{align}

We begin our proof by noting that for any set of orthogonal projectors and any unitarily invariant norm, we have a triangle(-type) inequality $\norm{\sum_k \Pi^{(k)} A \Pi^{(k)}}_1 \leq \norm{A}_1$.
Hence, we may write
\begin{align}
\norm{\arho_{AB}^{T_A}}_1 \geq & \norm{ \sum_{i,j} (\bar \Pi^{(i)}_A \otimes \Pi^{(j)}_B) \arho_{AB}^{T_A} (
\bar \Pi^{(i)}_A \otimes \Pi^{(j)}_B) }_1 
\nonumber \\
=&   \sum_{i,j} \norm{ (\bar \Pi^{(i)}_A \otimes \Pi^{(j)}_B) \arho_{AB}^{T_A} (\bar \Pi^{(i)}_A \otimes \Pi^{(j)}_B) }_1 
\nonumber  \\
& & \text{(Cont.)} \nonumber  
\end{align}
\begin{align}
=&  \sum_{i,j}  \norm{ \left[(\Pi^{(i)}_A \otimes \Pi^{(j)}_B) \arho_{AB} (\Pi^{(i)}_A\otimes \Pi^{(j)}_B) \right]^{T_A} }_1  \nonumber \\
=& \sum_{i,j} r_{ij} \norm{(\arho_{AB}^{(ij)})^{T_A}}_1,
  \label{eq:mono3_norm}
\end{align}
where  $\bar \Pi^{(i)}_A$ indicate an upside down projection operator (e.g., on the space $V^{\bar a_1\cdots \bar a_n}_{\bar a_1\cdots \bar a_n}$) and also satisfy $(\bar \Pi^{(i)}_A)^2=\bar \Pi^{(i)}_A$.
Moreover, in going from the second to the third line we use analog of Eq.~(\ref{eq:unitary_diagram}) for projection operators. Taking the logarithm of both sides and noting the fact that logarithm
is a concave function, we arrive at Eq.~(\ref{eq:mono3_logN}).

%%%%%%%%%%%%%%%%%%%%%%%%%%%%%%%%%%%%%%%%%%%%%%%%
%%%%%%%%%%%%%%%%%%%%%%%%%%%%%%%%%%%%%%%%%%%%%%%%

\section{Proof of $\norm{\arho^{T_A}}_1=\norm{\arho^{T_B}}_1$}

\label{app:proof Ta=Tb}
\renewcommand\theequation{D\arabic{equation}}

We prove this identity in two steps:
\begin{align}
    \label{eq:rTa_rTb}
    \norm{\arho^{T_A}}_1 = 
    \norm{(\arho^{T_A})^T}_1 = 
    \norm{\arho^{T_B}}_1
\end{align}
In the first identity, we need to show that the full transpose preserves the one norm, i.e.,
\begin{align}
    \label{eq:rT_norm}
    \norm{\arho^T}_1= \norm{\arho}_1.
\end{align}
In the second step, we must prove that
\begin{align}
    \label{eq:rTa_T}
    (\arho^{T_A})^T = \arho^{T_B}.
\end{align}

\noindent\textbf{Proof of Eq.~(\ref{eq:rT_norm}):}
The proof follows immediately from the diagrammatic approach,
\begin{align}
    \label{eq:trans_ab}
    \arho^{T}_{ab} &\defineas \sum_{f} \frac{p_{ab}^f}{\sqrt{d_a d_b d_f}}
    \begin{pspicture}[shift=-1.5](-1.9,-1)(1.2,2.3)
        \scriptsize
        % braid 
        \psline(-1,0)(0,1.5)
        \psline(-1.5,0)(0,2.25)
        \psline(1,1.5)(0,2.25)
        \psline[ArrowInside=->](-1.5,0)(-1,0.75)\rput(-1.55,0.3){$\bar b$}        \psline[ArrowInside=->](-1,0)(-0.66,0.51)\rput(-1,0.3){$\bar a$}
        \psline[ArrowInside=->](0,-0.75)(-0.5,0)\rput(-1.55,1.25){$\bar b$}
        \psline[border=2pt](-1.5,1.5)(0,-0.75)
        \psline(1,0)(0,-0.75)
        \psline[border=2pt](0,0)(-1,1.5)
        \psline[ArrowInside=->](0,0)(-0.33,0.5)\rput(-1,1.25){$\bar a$}
        \psline[ArrowInside=->](0,0)(0.5,0.5)\rput(0.4,0.2){$a$}
        \psline[ArrowInside=->](1,0)(0.5,0.5)\rput(1,0.25){$b$}
        % \rput(0.7,0.5){$\mu$}
        \psline[ArrowInside=->](0.5,0.5)(0.5,1)\rput(0.65,0.75){$f$}
        % \rput(0.7,1){$\mu$}
        \psline[ArrowInside=->](0.5,1)(0,1.5)\rput(0.35,1.35){$a$}
        \psline[ArrowInside=->,](0.5,1)(1,1.5)\rput(1,1.25){$b$}
    \end{pspicture} \\
    &= \sum_f \frac{p_{ab}^f \theta_f^\ast}{\sqrt{d_a d_b d_f}}
    \begin{pspicture}[shift=-0.6](-0.2,0)(1.2,1.5)
        \scriptsize
        \psline[ArrowInside=->](0,0)(0.5,0.5)\rput(0,0.25){$\bar a$}
        \psline[ArrowInside=->](1,0)(0.5,0.5)\rput(1,0.25){$\bar b$}
%        \rput(0.7,0.5){$\mu$}
        \psline[ArrowInside=->](0.5,0.5)(0.5,1)\rput(0.65,0.75){$\bar f$}
 %       \rput(0.7,1){$\mu$}
        \psline[ArrowInside=->](0.5,1)(0,1.5)\rput(0,1.25){$\bar a$}
        \psline[ArrowInside=->,](0.5,1)(1,1.5)\rput(1,1.25){$\bar b$}
    \end{pspicture}.
\end{align}
The second line is a special case of the identity (\ref{eq:rT_theta}).
Hence, we have
\begin{align}
    \arho^{T}_{ab} 
    \arho^{T\dag}_{ab}
    = \sum_f \left(\frac{p_{ab}^f}{\sqrt{d_a d_b d_f}}\right)^2
    \begin{pspicture}[shift=-0.6](-0.2,0)(1.2,1.5)
        \scriptsize
        \psline[ArrowInside=->](0,0)(0.5,0.5)\rput(0,0.25){$\bar a$}
        \psline[ArrowInside=->](1,0)(0.5,0.5)\rput(1,0.25){$\bar b$}
%        \rput(0.7,0.5){$\mu$}
        \psline[ArrowInside=->](0.5,0.5)(0.5,1)\rput(0.65,0.75){$\bar f$}
 %       \rput(0.7,1){$\mu$}
        \psline[ArrowInside=->](0.5,1)(0,1.5)\rput(0,1.25){$\bar a$}
        \psline[ArrowInside=->,](0.5,1)(1,1.5)\rput(1,1.25){$\bar b$}
    \end{pspicture}, 
\end{align}
which implies Eq.~(\ref{eq:rT_norm}).

\vspace{1ex}

\noindent\textbf{Proof of Eq.~(\ref{eq:rTa_T}):}
As before, we use diagrams to show this identity as follows,
\begin{align}
    (\arho^{T_A}_{ab})^T &= \sum_{f} \frac{p_{ab}^f}{\sqrt{d_a d_b d_f}}
    \begin{pspicture}[shift=-1.9](-2.3,-1.3)(1.3,2.4)
        \scriptsize
        % braid
        \psline(0,0)(-0.8,1.3)
        \psline(-0.8,1.3)(-1.5,0)
        \psline[border=2pt](-0.8,0.2)(0,1.5)
        \psline[ArrowInside=->](0,0)(-0.33,0.54)
        % transpose
        \psline(1,1.5)(-0.6,2.5)
        \psline(-0.6,2.5)(-2,0)\rput(-2.1,0.1){$\bar b$}
        \psline[border=2pt](-0.6,-1)(-2,1.5)\rput(-2.1,1.4){$\bar b$}
        \psline(1,0)(-0.6,-1)
        \psline[border=2pt](-0.8,0.2)(-1.5,1.5)\rput(-1.6,1.4){$a$}
        \psline[ArrowInside=->](-0.8,0.2)(-0.5,0.69)\rput(-1.6,0.1){$a$}
        \psline[ArrowInside=->](0,0)(0.5,0.5)\rput(0.4,0.2){$a$}
        \psline[ArrowInside=->](1,0)(0.5,0.5)\rput(1,0.25){$b$}
        \psline[ArrowInside=->](0.5,0.5)(0.5,1)\rput(0.65,0.75){$f$}
        % \rput(0.7,1){$\mu$}
        \psline[ArrowInside=->](0.5,1)(0,1.5)\rput(0.35,1.35){$a$}
        \psline[ArrowInside=->,](0.5,1)(1,1.5)\rput(1,1.25){$b$}
    \end{pspicture}\\
    &= \sum_{f} \frac{p_{ab}^f}{\sqrt{d_a d_b d_f}}
    \begin{pspicture}[shift=-0.6](-1.2,0)(1.2,1.5)
        \scriptsize
        % braid
        \psline(-1,0)(0,1.5)
        \psline[border=2pt](0,0)(-1,1.5)
        \psline[ArrowInside=->](0,0)(-0.33,0.5)\rput(-1,1.2){$\bar b$}
        \psline[ArrowInside=->](-1,0)(-0.66,0.5)\rput(-1,0.3){$\bar b$}
        \psline[ArrowInside=->](0,0)(0.5,0.5)\rput(0.4,0.2){$b$}
        \psline[ArrowInside=->](1,0)(0.5,0.5)\rput(1,0.25){$a$}
        % \rput(0.7,0.5){$\mu$}
        \psline[ArrowInside=->](0.5,0.5)(0.5,1)\rput(0.65,0.75){$f$}
        % \rput(0.7,1){$\mu$}
        \psline[ArrowInside=->](0.5,1)(0,1.5)\rput(0.35,1.35){$b$}
        \psline[ArrowInside=->,](0.5,1)(1,1.5)\rput(1,1.25){$a$}
    \end{pspicture}\\
    &=\arho_{ab}^{T_B},
\end{align}
where we use the fact that the two half braids from top and bottom of the diagram cancel each other, since $|R^{ba}_f|^2=1$.
This completes the proof of Eq.~(\ref{eq:rTa_rTb}).

%%%%%%%%%%%%%%%%%%%%%%%%%%%%%%%%%%%%%%%%%%%%%%%%%%%

\section{\texorpdfstring{$F$ and $R$}{F and R} symbols}
\label{app:Fsymbols}
\renewcommand\theequation{E\arabic{equation}}

Here, we provide the anyon data of $su(2)_k$ and $su(3)_3$ categories 
which we use to calculate the logarithmic negativity in Secs.~\ref{sec:examples-su2} and \ref{sec:examples-su3} of main text.

\subsection{\texorpdfstring{$su(2)_k$}{su(2) lvl k}}

The corresponding $F$-symbols are given by
\begin{align}
    \left[F^{j_1,j_2,j_3}_j\right]_{j_{12},j_{23}}
    &= (-1)^{j_1+j_2+j_3+j}
    \sqrt{[2j_{12}+1]_q [2j_{23}+1]_q}\nonumber \\
    &\quad\times \left\{
    \begin{matrix}
    j_1 & j_2 & j_{12} \\
    j_3 & j & j_{23}
    \end{matrix}
    \right\}_q,
\end{align}
where $q=e^{i\frac{2\pi}{k+2}}$, $[n]_q=\frac{q^{n/2}-q^{-n/2}}{q^{1/2}-q^{-1/2}}$ is a $q$-deformed number and $\{ . \}_q$ is $q$-deformed version of $su(2)$ $6j$-symbols.
The $R$-symbols are
\begin{align}
    R^{j_1,j_2}_j = (-1)^{j-j_1-j_2}
    q^{\frac{1}{2}[j(j+1)-j_1(j_1+1)-j_2(j_2+1)]},
\end{align}

It is worth recalling that $su(2)_2$ with $\{0,\frac{1}{2},1\}$ spins and $su(2)_3$ with $\{0,1\}$ spins are in one-to-one correspondence with the $\nu=3$ Ising and Fibonacci anyons, respectively.

\subsection{Subtheory of \texorpdfstring{$su(3)_3$}{su(3) lvl 3}}

The non-trivial $F$-symbols of this category (based on Ref.~\cite{Ardonne_2010}) are given by
\begin{align}
    F^{1,8,8}_8 &= F^{8,1,8}_8 = F^{8,8,1}_8= F^{8,8,8}_1=
    \begin{pmatrix}
    1 & 0 \\ 0 & 1
    \end{pmatrix},  
\end{align}
\begin{align}
    F^{8,8,8}_{10} &= F^{8,8,10}_8 = F^{8,\overline{10},8}_8= F^{10,8,8}_8= \begin{pmatrix}
    -\frac{1}{2} & -\frac{\sqrt{3}}{2} \\ \frac{\sqrt{3}}{2} & -\frac{1}{2}
    \end{pmatrix},  \\
    F^{8,8,8}_{\overline{10}} &= F^{8,8,\overline{10}}_8 = F^{8,10,8}_8= F^{\overline{10},8,8}_8= \begin{pmatrix}
    -\frac{1}{2} & \frac{\sqrt{3}}{2} \\ -\frac{\sqrt{3}}{2} & -\frac{1}{2}
    \end{pmatrix},
\end{align}
where $2\times 2$ matrices are due to the two-fold multiplicity of fusing two anyon-$8$'s.
We also have
\begin{align}
    F^{888}_8 &= \renewcommand{\arraystretch}{1.2} \begin{pmatrix}
    \frac{1}{3} & \frac{1}{\sqrt{3}} & 0 & 0 &
    \frac{1}{\sqrt{3}} & -\frac{1}{3} & -\frac{1}{3} \\
    \frac{1}{\sqrt{3}} & -\frac{1}{2} & 0 & 0 &
    \frac{1}{2} & \frac{1}{\sqrt{12}} & \frac{1}{\sqrt{12}} \\
    0 & 0 & \frac{1}{2} & \frac{1}{2} &
    0 & \frac{1}{2} & -\frac{1}{2} \\
    0 & 0 & \frac{1}{2} & \frac{1}{2} &
    0 & -\frac{1}{2} & \frac{1}{2} \\
    \frac{1}{\sqrt{3}} & \frac{1}{2} & 0 & 0 &
    -\frac{1}{2} & \frac{1}{\sqrt{12}} & \frac{1}{\sqrt{12}} \\
    -\frac{1}{3} & \frac{1}{\sqrt{12}} & -\frac{1}{2} & \frac{1}{2} &
      \frac{1}{\sqrt{12}} & \frac{1}{3} & \frac{1}{3} \\
    -\frac{1}{3} & \frac{1}{\sqrt{12}} & \frac{1}{2} & -\frac{1}{2} &
      \frac{1}{\sqrt{12}} & \frac{1}{3} & \frac{1}{3}
    \end{pmatrix},
\end{align}
where the matrix $[F^{8,8,8}_8]_{(e,\alpha,\beta),(f,\mu,\nu)}$ matrix is written in the basis such that
the first, sixth, and seventh rows/columns correspond to $e=1,10,\overline{10}$, respectively. The second to fifth rows/columns correspond to $e=f=8$. In this case, there are two vertices with three  anyon $8$ lines. The second and fifth row correspond to the cases in which we take the vertices to be the same $\alpha=\beta$ and $\mu=\nu$, while the third and fourth row correspond to the off-diagonal cases $\alpha\neq \beta$ and $\mu\neq \nu$. The non-trivial $R$-symbols read 
\begin{align}\begin{aligned}
    R^{1x}_x &= 1, \\
    R^{88}_1 &= -1, \qquad R^{88}_{10}=R^{88}_{\overline{10}}=-1,\\
    R^{88}_8 &=
    \begin{pmatrix}
    -i & 0 \\ 0 & i
    \end{pmatrix} .
\end{aligned}\end{align}

\emph{Derivation of negativity$-$.} 
Partial transpose of density matrix (\ref{eq:su3}) can be written in the form of Eq.~(\ref{eq:pt_dimer}) where $[M^c]$'s are given by
\begin{align} \begin{split}
    [M^1] &= 2p-1,\\
    [M^{10}] &= \frac{i}{2}(2p-1)+ i \sqrt{3} q_r,\\
    [M^{\overline{10}}] &= \frac{i}{2}(2p-1)- i \sqrt{3} q_r,\\
    [M^8] &=\begin{pmatrix}
    \frac{i}{2} & q_i \\
    q_i & -\frac{i}{2}
    \end{pmatrix}.
\end{split} \end{align}
Note that $[M^c]$'s are scalar except for $[M^8]$ which has a fusion multiplicity.
Plugging these values in for Eq.~(\ref{eq:ab_LN_mu}) leads to \eqref{eq:LN_su3}.

%%%%%%%%%%%%%%%%%%%%%%%%%%%%%%%%%%%%%%%%%%%%%%%%%%%%%%
%%%%%%%%%%%%%%%%%%%%%%%%%%%%%%%%%%%%%%%%%%%%%%%%%%%%%%

\bibliography{refs}

\end{document}